\documentclass[english, a4paper]{amsart}
\usepackage[T1]{fontenc}
\usepackage[latin9]{inputenc}
\usepackage{xcolor}
\usepackage{dsfont}
\usepackage{amstext}
\usepackage{amsthm}

\usepackage{amssymb}
\usepackage{graphicx}
\usepackage{tablefootnote}
\usepackage{esint}
\usepackage{soul}
\usepackage{multirow}
\usepackage{float}

\let\oldtocsection=\tocsection  % for toc
\let\oldtocsubsection=\tocsubsection % for toc
\renewcommand{\tocsection}[2]{\hspace{0em}\oldtocsection{#1}{#2}} % for toc
\renewcommand{\tocsubsection}[2]{\hspace{2em}\oldtocsubsection{#1}{#2}} % for toc
\newtheorem{innercustomgeneric}{\customgenericname}
\providecommand{\customgenericname}{}
\newcommand{\newcustomtheorem}[2]{%
  \newenvironment{#1}[1]
  {%
   \renewcommand\customgenericname{#2}%
   \renewcommand\theinnercustomgeneric{##1}%
   \innercustomgeneric
  }
  {\endinnercustomgeneric}
}

\newcustomtheorem{customprop}{Proposition}
\newcustomtheorem{customremark}{Remark}

\makeatletter

\providecommand{\tabularnewline}{\\}

\numberwithin{equation}{section}
\numberwithin{figure}{section}
\theoremstyle{plain}

\theoremstyle{plain}

\theoremstyle{remark}

\makeatother

\usepackage{babel}
\providecommand{\propositionname}{Proposition}
\providecommand{\remarkname}{Remark}
\providecommand{\theoremname}{Theorem}

\title[Tools for goodness-of-fit validation]{A new set of tools for goodness-of-fit validation}

\author[G.R. Ducharme and T. Ledwina]{by Gilles~R. DUCHARME \\\\ {\it IMAG, Univ Montpellier, CNRS, Montpellier, France \\ \lowercase{e.mail : gilles.ducharme@umontpellier.fr}} \\\\ and Teresa LEDWINA \\\\ {\it Institute of Mathematics, Polish Academy of Sciences} \\{\it ul. Kopernika 18, 51-617 Wroc{\l}aw, Poland \\  \lowercase{e.mail : ledwina@impan.pl} } }


\begin{document}

\begin{abstract}
We introduce two new tools to assess the validity of statistical distributions. These are based on components derived from a new statistical quantity, 
the {\it comparison curve}, which can provide a detailed appraisal of validity. The first tool is a graphical representation of these components on a {\it bar plot} (B-plot). The knowledge such gained could suggest an existing {\it goodness-of-fit test} to supplement this assessment with a control of the type I error. Otherwise, an adaptive test may be preferable and the second tool is the combination of these components  to produce a powerful $\chi^2$-type {\it goodness-of-fit test}. Because the number of these components can be  large, we introduce a new selection rule to decide on their number. 
In a simulation, our goodness-of-fit tests are powerwise competitive with the best solutions recommended. Practical examples show how to use these tools to derive principled information about where the model departs from the data.

\end{abstract}

\maketitle

\vspace{-1.cm}

\footnotesize{\tableofcontents}  % for the table of content

%\begin{keywords}
%{Chi-square test; Data driven test; Diagnostic component; Graphical inference; Model validation; Score statistic; Selection rule; Smooth test.}
%\end{keywords}

\normalsize

\section{Introduction\label{sec:Introduction}}

Let $X$ be a random variable with unknown cumulative distribution
function (CDF) $F(\cdot).$ Statistical models are entertained approximations
to $F(\cdot)$ which serve to produce inferential statements about
the behaviour of $X$. Constructing a good approximation is an iterative
process where at any given step, a contemplated model based on previously
acquired knowledge is assessed by
confrontation with data. When the current proposal is invalidated,
its defects must be learned to explore a better model at the next
iteration. When a model is tentatively validated, useful inference can
be drawn by exploiting its characteristics, allowing the accumulation
of subject matter knowledge.

In the present work, the entertained statistical model for $X$ is the CDF $\text{\ensuremath{F_{0}}(\ensuremath{\cdot}\,;\ensuremath{~\beta})}$
where the parameter $\beta$ may be unknown. The data is
a sample of independent copies $X_{1},\ldots,X_{n}$ of $X.$

Two main routes exist for statistical model validation. A first one
focusses on graphical representations, such as PP (percentile-percentile)
or QQ (quantile-quantile) plots (Thas, 2010, Section 3.2). When the model is valid,
these plots should closely follow the 45-degree line through the origin.
Deviations can provide insights about where the data do not conform
to the entertained model. These visual appreciations can be supplemented
with confidence regions about such representations 
(Aldor-Noiman et al., 2013; 
Gan, Koehler \& Thompson, 1991) or test statistics measuring
departure from this straight line (Gan \& Koehler, 1990) to 
control  the type I error (i.e. falsely considering a model is
invalid). This route can
clearly be a cog in the modelling process.

A second route focusses mainly on error risks by testing, via formal
goodness-of-fit (GoF) procedures, the null hypothesis that the model
holds. A large number of test statistics have been derived for such
problems (for testing the GoF to the Gaussian distribution, Arnastauskait\'e, Ruzgas \& Braz\'enas (2021) list 40 such tests) which allow controlling the type
I error risk. Regarding the type II error (i.e. not rejecting an invalid
model and thus stopping prematurely the modelling process), an enormous
amount of work has been accomplished, both theoretically and empirically,
to understand the respective power of the various proposals and to
derive a generally good solution for specific problems. In particular,
regarding the Gaussian distribution again, a long series of simulation
experiments have been conducted (see Arnastauskait\'e, Ruzgas \& Braz\'enas, 2021
and references therein) to characterize the effectiveness of popular
proposals. A first drawback is that it is not easy to decide,  in view of the data at hand, upon
an appropriate GoF test among this plethora of solutions. Another drawback is that when the chosen test
rejects, the user is often left with little information about the
defects of the model. This makes it difficult to pursue the modelling. 

A few exceptions are the well-known Pearson $\chi^{2}$-test and
the smooth test introduced by Neyman (1937), to which
sets of components can be associated. Each component reacts to specific
departures and if these can be discerned, their inspection can help
a user gain some insights about where the model is at fault. 
Below, we discuss some problems arising with these tests when $\beta$ is known.
Both  approaches have been extended to unknown $\beta$ but then
more serious difficulties occur.

There are two main problems with such tests. The
first one arises from the standard order of the argument leading to
these components, which is first to get from external
considerations  a GoF test, then try to extract meaningful components. For Pearson's
$\chi^{2}$-test, Anderson (1994) has made such an attempt but Boero, Smith \& Wallis
(2004a) have shown that his approach was not completely successful;
for similar efforts, see Voinov (2010) and references therein. When $\beta$ is known, 
the components of the classical Pearson's $\chi^2$-square test
are not easy to interpret as they heavily depend on the selected partition. Regarding
Neyman's smooth tests, meaning depends on the orthogonal system used
in the test. With classical orthogonal polynomials, the first few
components will typically be associated with central moments (Thas, 2010, p. 84),
but beyond the third (skewness) they become difficult to relate
to telling departures.  

The second problem concerns the number of components
to be used in the test: too few or too many negatively affect the power of the test. 
For some historical notes and recommendations about choosing the partitions,
see  Boero, Smith \& Wallis (2004b) and Rolke $\&$ Gongora (2021). Bogdan (1995) and Inglot $\&$ Janic-Wr\'oblewska (2003) contain 
some useful proposals regarding data driven selections of the partition in classical $\chi^2$-test. Some
data driven GoF tests based on partitions have also been derived in  Section~5.5 of Thas (2001).
In turn, Ledwina (1994) has proposed an effective way of selecting the 
number of components in the classical Neyman's smooth test.

In the present work, we try to solve these problems by introducing
easily interpretable components, different from those in the original $\chi^2$-test and several variants
of Neyman's test, which can be graphically depicted \textcolor{black}{ on what we call a {\it bar plot} (B-plot)}
to give a detailed appraisal of the validity of a model.  This may help in selecting an appropriate GoF test 
to assess the global fit between the data and the model. When, likely, none naturally emerges, the components can be 
combined to produce a powerful new data driven $\chi^{2}$-type GoF test, \textcolor{black}{ (see Section 2.4 for an explanation of the label ``-type''),} supplementing
the visual assessment with a control of the type I error.  \textcolor{black}{ When the chosen test rejects, acceptance regions for subsets of components under the null model
can be plotted and analyzed to sharpen the insight gained from the B-plot about where and why departures seem to occur.}

We start by introducing a function, referred to as the {\it comparison curve}
(CC), which is related to existing statistical objects such
as PP and QQ plots. Its evaluation yields components whose statistical properties
offer richer insights, when depicted on the B-plot,
than these plots. In particular,
this first new tool allows gaining some ideas about where (in terms of ranges of quantiles of the model $F_{0}(\cdot\,;\,\beta)$)
and to which extent the data contradict the model. These components are then shown to be estimated successive Fourier coefficients
of the {\it comparison density} (see Section~2.3).
The number of such components can in principle be as large as one
chooses. Hence a second important task is to derive a way of selecting,
in a data driven fashion, their number. We introduce a new selection
rule to decide on a proper number of components to include in our second tool, a {\it data driven
$\chi^{2}$-type GoF test statistic}. Finally, we show how the B-plot can be supplemented with acceptance regions for subsets of components to provide richer indications regarding the compatibility of the data with the null model in some regions of interest.

We first  consider the context where the null model $\text{\ensuremath{F_{0}}(\ensuremath{\cdot}\,;\,\ensuremath{\beta})}$
is entirely specified, i.e. $\beta$ is known. A
carefully balanced simulation experiment shows that our procedure
competes with some best tests in this context. Then we move to the context where the parameter $\beta$ in $\text{\ensuremath{F_{0}}(\ensuremath{\cdot}\,;\,\ensuremath{\beta})}$
must be estimated. Particular attention is given to the location-scale
model, i.e. $F_{0}\big((x-\beta_{1})/\beta_{2}\big)$, and to the important
sub-case of a Gaussian model. Again the results of a balanced simulation
experiment show that our procedure competes with some best tests for this
set-up. In both contexts, we apply our tools to real data to
show how useful insights can be derived. The appendices contain details about  more examples showing the useful  information that can derive from the tools of the paper, some practical recommendations regarding the application of our test strategies, the results of a simulation study regarding the power of the test statistic of Section 3 along with some more general discussion and the proofs of various technical results.

% ############## Section 2.0 #########################

\section{The case of a simple null hypothesis\label{sec: 2 Case 0}}

\subsection*{2.1.~~Comparison curve (CC) and B-plot\label{subsec:The-comparison-curve Section 2.1}}

Let $X_{1},\ldots,X_{n}$ be a sample of i.i.d. observations from an unknown
continuous CDF $F(\cdot)$. We start by considering the case where
the parameter $\beta$ in the continuous model $\text{\ensuremath{F_{0}}(\ensuremath{\cdot}\,;\,\ensuremath{\beta})}$
is known and write for simplicity $F_{0}(\cdot)$ for this CDF. The simple null
hypothesis of interest is $\mathbb{H}_{0}:F(\cdot)=F_{0}(\cdot)$. 

Consider the random variable $Z=F_{0}(X)$. By the {\it probability
integral transformation}, when $X\sim F(\cdot)$, $Z$ has CDF $H(p)=F(F_{0}^{-1}(p))$,
$p\in(0,1)$, which is referred to by Parzen (2004) as the {\it comparison
CDF}, because when some auxiliary random variable $X_{0}\sim F_{0}(\cdot)$,
then $H(\cdot)$ is the CDF of $X$ expressed on a scale in which
$X_{0}\sim U(0,1).$ $H(\cdot)$ is also referred to in the literature
as the {\it relative distribution} (Handcock \& Morris, 1999, Chapter 2, p. 21) as
$Z$ measures the relative ranks of $X$ compared to $X_{0}\sim F_{0}(\cdot)$.
Such relative ranks are also known as the {\it grade transformation}
following a statistical tradition that goes back to Galton; cf. Kendall
\& Buckland (1957, p. 121). The function $H(\cdot)$ is also the population
version of the PP plot of $F(\cdot)$ against $F_{0}(\cdot)$ which, in this context, is sometimes called  the reference distribution.

The approach of the present work is based on a standardized version
of the {\it comparison CDF}, which we call the {\it comparison curve}
(CC) and define as 
\begin{alignat}{1}
\textnormal{CC}(p) & =\frac{p-F(F_{0}^{-1}(p))}{(p(1-p))^{1/2}},\quad p\in(0,1).\label{eq:CC(p) curve)}  % eq (2)
\end{alignat}
When $\mathbb{H}_{0}$ holds$,\textnormal{CC}(\cdot)\equiv0$ and otherwise captures
weighted vertical discrepancies between the population PP plot and
the 45-degree line.  As with PP plots, $\textnormal{CC}(\cdot)$ is invariant under
\textcolor{black}{strictly increasing and continuous transformations} of the  scale of measurement. But in contrast with PP plots which are
always 0 as $p\rightarrow0$, and 1 as $p\rightarrow1$, $\textnormal{CC}(\cdot)$ can be unbounded at the boundaries, 
see  the Lehmann contamination and the Anderson kurtotic alternatives in Figure \ref{fig:CC(curves) Case 0 and 3}.
\textcolor{black}{As a result,} due to the meaningful weighting in (\ref{eq:CC(p) curve)}), $\textnormal{CC}(\cdot)$ can better exhibit differences between $F(\cdot)$ and
$F_0(\cdot)$ appearing in tails.  The equality  $\textnormal{CC}(p) = 0$ for all $p\in (0,1)$ is equivalent to $F(x) = F_0(x)$ for all $x$, while $\textnormal{CC}(p)\geq0$ for all $p$ is equivalent
to  $F(\cdot)$ being stochastically larger than $F_0(\cdot)$. These properties are well known with PP plots. However, we can say more on both PP and CC plots in terms of probability mass allocation between 
$F(\cdot)$ and $F_0(\cdot)$ in relation with their stochastic ordering. Namely,  if there is only one point $p_0 \in (0,1)$ such that $\textnormal{CC}(p_0)=0$, then $F^{-1}(p_0)=F_{0}^{-1}(p_0)$ and consequently
the set $(-\infty,  F_{0}^{-1}(p_0)]$  has the same probability under both $F(\cdot)$ and $F_0(\cdot)$.  Obviously, the same conclusion  holds
for the set $( F_{0}^{-1}(p_0), +\infty)$. The relation $\textnormal{CC}(p)>0$ on $(0,p_0)$ defines the region where $F_0(\cdot)> F(\cdot)$. Hence, 
when restricted to this interval, observations generated from the conditional distribution of $F(\cdot)$ are
stochastically larger than under the respective conditional variant of $F_{0}(\cdot)$.  Otherwise stated,  the probability mass associated with $F(\cdot)$  accumulates 
more intensively toward $F_0^{-1}(p_0)$ than the mass of $F_0(\cdot)$.   In terms of quantiles, we get 
$F_{0}^{-1}(\cdot)<F^{-1}(\cdot)$ and  the quantiles of $F(\cdot)$ are more concentrated toward the $p_{0}$-quantile of the reference CDF $F_{0}(\cdot)$ than those
of $F_0(\cdot)$ itself.  The reverse holds when $\textnormal{CC}(\cdot)<0$ on $(0, p_0)$. The magnitude of $\textnormal{CC}(\cdot)$ reflects the rate at which the mass allocation between the two CDFs changes. If there are more than one point $p$ such that $\textnormal{CC}(p) =0$, these interpretations apply to each resulting region in $(0, 1)$. The above, along with the comments in Section 3.3, essentially strengthens the interpretation of PP plots  discussed in Thas (2010), Sections 7.6 and 8.1.1.2, and lead to the view that  $\textnormal{CC}$ plots can be seen as upgraded variants of PP plots.

Replacing in (\ref{eq:CC(p) curve)}) the unknown $F(\cdot)$ by $\hat{F}_{n}(x)=n^{-1}\sum_{i=1}^{n}I(X_{i}\leq x)$,
where $I(\omega)$ is the indicator function of event $\omega$,
leads to the {\it empirical \textnormal{CC}} 
\begin{align}
\widehat{ \textnormal{CC}}(p) & =\frac{p-\hat{F}_{n}(F_{0}^{-1}(p))}{(p(1-p))^{1/2}},\quad p\in(0,1).\label{eq:Empirical CC case 0}
\end{align}
Formally, $\widehat{\textnormal{CC}}(p)$ is a consistent estimator of $\textnormal{CC}(p)$ in the sense that, for any $\epsilon \in (0,1)$,
$\sup_{\epsilon \leq p \leq 1-\epsilon} \bigl|\widehat{\textnormal{CC}}(p)-\textnormal{CC}(p) \bigr| \rightarrow 0$ in probability.
More importantly,  $n^{1/2}\,\widehat{\textnormal{CC}}(p)$ is asymptotically
$N(0,1)$ under $\mathbb{H}_{0}$ for each $p$. Thus, in contrast
with empirical PP and QQ plots, $n^{1/2}\,\widehat{\textnormal{CC}}(\cdot)$
captures discrepancies between the postulated model under $\mathbb{H}_{0}$ and the data with
equal precision over the whole range of $p$. 

Evaluating $n^{1/2}\,\widehat{\textnormal{CC}}(\cdot)$
at points on a grid in $(0,1)$ and representing these as bars over
the grid points yields a {\it bar plot} (B-plot), as introduced by Ledwina
and Wy{\l}upek (2012a, b) in a related problem.
Obviously, the $n^{1/2}\,\widehat{\textnormal{CC}}(\cdot)$ are noisy but
being correlated, their visual inspection can allow to
approximately identify regions where the null model puts more probability
mass, via a clustering of its quantiles, than the data seems to suggest,
and reversely.

\begin{figure}[H]
\centering
\begin{centering}
\includegraphics[width = \linewidth]{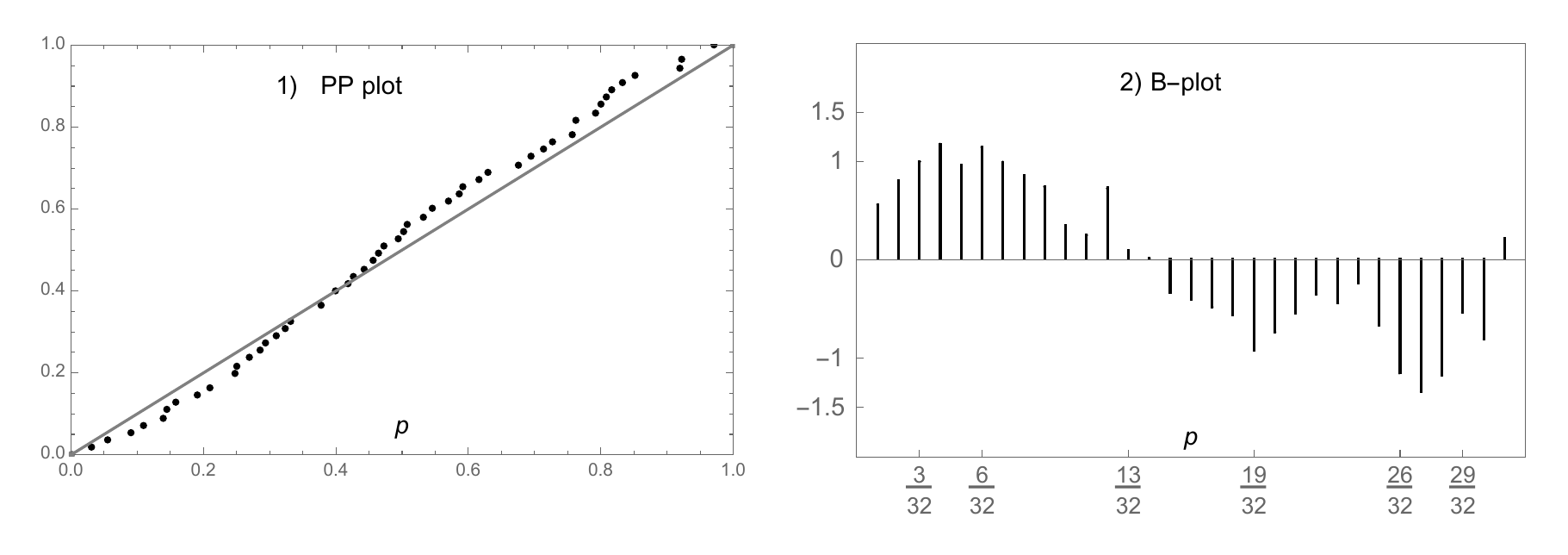}
\par\end{centering}
\caption{\label{fig:miling baby data set} Graphical representations for the
smiling baby data ($n=55)$ in Bhattacharjee \& Mukhopadhyay (2013).
Panel 1) : PP plot against the $U[0,1]$ distribution (45-degree
 solid line) ; Panel 2): B-plot of the $n^{1/2}\,\widehat{\textnormal{CC}}(\cdot)$
of eq. (\ref{eq:Empirical CC case 0}) evaluated over the grid $p\in\{1/32,\ldots,31/32\}$.}
\end{figure}

As an example of the usefulness of a $\textnormal{CC}(\cdot)$, consider the
smiling baby data set (Bhattacharjee \& Mukhopadhyay, 2013). The data
($n=55)$ are the smiling times (in seconds) of an eight-week-old
baby. According to various authors, the data could realistically be
uniformly distributed over the interval {[}0, $\theta${]}. Here,
we take $\theta=23.5$, a value close to the estimators investigated
by Bhattacharjee \& Mukhopadhyay (2013) and transform the data onto
$[0,1]$. Panel 1) of Figure \ref{fig:miling baby data set} shows
the empirical PP plot against the reference $U[0,1]$ distribution
(the 45-degree line) while Panel 2) represents the B-plot of
$n^{1/2}\,\widehat{\textnormal{CC}}(p)$ for $p\in\{1/32,\ldots,31/32\}$. 
Inspection of these bars shows a coincidence of the null and empirical
quantiles in the neighbourhood of $p=0.45\;(\simeq10.6$ in the original
units), slightly to the left of the median under $\mathbb{H}_{0}$.
The shape of the sets of positive bars to the left and negative bars to the right of $p \simeq 0.45$ suggests that the central quantiles
of the true distribution could be more clustered about this point  
than those of the null uniform. Thus the true distribution is perhaps less dispersed than the uniform
and slightly shifted toward 0. Similar insight can be derived
from the empirical PP plot (Panel 1) and QQ plot (not shown) of this
data set. However, by making use of the null expectation and approximate
homoscedasticity of $\widehat{\textnormal{CC}}(\cdot)$ under $\mathbb{H}_{0}$,
the B-plot allows an enhanced appraisal of these main features
of the data. In particular, this B-plot allows seeing for which
quantiles of the null CDF $F_{0}(\cdot)$, the discrepancy with the
data is unexpectedly large (positive or negative). The approach is developed in Section~3.3 for the null and composite case and
discussed for the present case in Section~3.5.4, where we revisit this data set.

Such visual insights about the discordance of data with a null model
is interesting, but should be supported by a GoF test for error control. 
Here we exploit the asymptotic behaviour of $n^{1/2}\,\widehat{\textnormal{CC}}(\cdot)$
under $\mathbb{H}_{0}$ to obtain inferential statements about the overall
validity of the model. Hence, we now consider the problem of creating
a GoF test based on the empirical CC. Here, we proceed in a
simple and traditional way by considering $\widehat{\textnormal{CC}}(p)$
evaluated at points $p$ in a finite set associated with a B-plot of interest. These points are described
in the next subsection. Then, from the values of the related bars, we build a $\chi^{2}$-type test statistic.
Finally, we introduce a selection rule to decide about the most useful
subset of  points, which is a highly non-trivial problem in the
case of statistics of the present type.

%% ############    2.2 : Nested partition of (0,1) and projected Haar functions.    ################

\subsection*{2.2.~~Nested partition of (0,1) and projected Haar function $\{h_{s,j}(\cdot)\}$\label{subsec:Nested-partition-of And Haar function Section 2.2}}

Let  $a_{s,k}={(2k-1)}/{2^{s+1}},  (s=0,1,\ldots;\;\;k=1,2,\ldots,2^{s})$. Associated with a sequence of sample sizes $n$, let $S(n)$
be a user-defined increasing sequence of integers. With $s$ ranging
in $\{0,1,\ldots,S(n)\}$ and $k\in\{1,2,\ldots,$2$^{s}$$\}$,
introduce in turn the sequence of nested sets of points in $(0,1)$
corresponding to those $a_{s,k}$'s with $\{s=0\},$ $\{s=0,1\},$
$\{s=0,1,2\},\ldots$, sorted from smallest to largest to create the
increasingly finer sets of points 
\begin{align*}
\ensuremath{\{p_{0,1}\},\;\{p_{1,1},p_{1,2},p_{1,3}\},\ldots,\{p_{s,1},\ldots,p_{s,d(s)}\},\ldots,\quad\;\;d(s)=2^{s+1}-1}.
\end{align*}
For example, if $S(n)=2$ the nested sets of ordered points
are $\{4/8\},\{2/8,4/8,6/8\}$ and $\{1/8,2/8,3/8,4/8,5/8,6/8,7/8\}$.
Also define $\ensuremath{{\mathbb{D}}(n)=\{d(s):s=0,\ldots,S(n)\}$ and $D(n)=d(S(n))}.$
%\begin{align*}
%\ensuremath{{\mathbb{D}}(n)=\{d(s):s=0,\ldots,S(n)\},\quad D(n)=d(S(n))}.
%\end{align*}

Now for a given $s$ and corresponding $d(s)$, introduce the $d(s)-$dimensional
vector of functions $(h_{s,1}(p),\ldots,h_{s,d(s)}(p))$ with $0\leq p\leq1$,
where for $j\in \{1,\ldots,d(s)\}$
\begin{align*}
h_{s,j}(p) & =-\left(\frac{1-p_{s,j}}{p_{s,j}}\right)^{1/2} \times I(0\leq p\leq p_{s,j})+\left(\frac{p_{s,j}}{1-p_{s,j}}\right)^{1/2}\times I(p_{s,j}<p\leq1) \\
&=\frac{p_{s,j}-I(0\leq p\leq p_{s,j})}{(p_{s,j}(1-p_{s,j}))^{1/2}}.
\end{align*}

These functions arise as normalized orthogonal projections of the
Haar functions onto the cone of nondecreasing functions (cf. Ledwina
\& Wy{\l}upek, 2012b) and constitute the building blocks of our
tools. Obviously, the functions in this system are
normalized but not orthogonal. The explicit form of the inner product
matrix of the $h_{s,j}(\cdot)$ and its inverse have been derived
in Ledwina \& Wy{\l}upek (2012b) under $\mathbb{H}_{0}$. Note that
Pearson's $\chi^{2}$ is also related to a set of  points $0=\pi_{0}<\pi_{1}<\ldots<\pi_{k}=1$
defining a normalized but not orthogonal system of functions given
by $l_{j}(p)={\{I(\pi_{j-1}<p<\pi_{j})-(\pi_{j}-\pi_{j-1})}\}/{(\pi_{j}-\pi_{j-1})^{1/2}}$. \textcolor{black}{ Indeed, Pearson's $\chi^2 = \sum_{j=1}^k \big(\sum_{i=1}^{n}\ell_j(X_i)\big)^2$}.
Here, a single $l_{j}(\cdot)$ corresponds to two neighbouring points, so Pearson's system is naturally adapted to histograms. This
is to be contrasted with the system $\{h_{s,j}(\cdot)\}$,
where each point $p_{s,j}$ corresponds to the single function $h_{s,j}(\cdot$)
and is thus adapted to CDFs. 

%% ############    2.3 : Fourier coefficients of the comparison density  in the system    ################

%\subsection{Fourier coefficients of the comparison density  in the system $\{h_{s,j}(\cdot)\}$\label{subsec:Section Fourier-coefficients}}

\subsection*{2.3.~~Fourier coefficient of the comparison density  in the system $\{h_{s,j}(\cdot)\}$\label{subsec:Section Fourier-coefficients}}

Write $f_{0}(\cdot)$ and $f(\cdot)$ for the densities of $F_{0}(\cdot)$
and $F(\cdot)$ respectively. Assume further that $f_{0}(x)=0\Longrightarrow$
$f(x)=0$. Then the function $H(\cdot)=F(F_{0}^{-1}(\cdot))$ satisfies $H(0)=0,$ $H(1)=1$
and possesses a density, called   the {\it comparison density} (Parzen, 2004) or the {\it relative density} (Handcock
\& Morris, 1999, Chapter 2, p. 22) given by $\kappa(p) = f(F_{0}^{-1}(p))/{f_{0}(F_{0}^{-1}(p))},~p\in(0,1)$.
Obviously, $\kappa(\cdot)\equiv1$ if and only if $\mathbb{H}_{0}$ holds. Now,
consider the Fourier coefficients (FC) of $\kappa(\cdot)$ in the
system $\{h_{s,j}(\cdot)\}.$ The $(s,j)$-th Fourier coefficient,
noted $\gamma_{s,j}$, takes the form 
\begin{align}
\gamma_{s,j}=\gamma_{s,j}(p_{s,j}) & =\int_{0}^{1}\kappa(p)\,h_{s,j}(p)\,dp = \frac{p_{s,j}-F(F_{0}^{-1}(p_{s,j}))}{(p_{s,j}(1-p_{s,j}))^{1/2}}.\label{eq:Fourier coefficient Case 0}
\end{align}
Then,  $\mathbb{H}_{0}$ can be equivalently reformulated as $\gamma_{s,j}=0, (s = 0,1,\ldots; \:  j=1,\ldots,d(s)).$

Expression (\ref{eq:Fourier coefficient Case 0}) leads to the empirical
FC : $\hat{\gamma}_{s,j}$ $=n^{-1}\sum_{i=1}^{n}h_{s,j}(F_{0}(X_{i}))$.
A little algebra shows that $\gamma_{s,j}=\textnormal{CC}(p_{s,j})$ and $\hat{\gamma}_{s,j}=\widehat{\textnormal{CC}}(p_{s,j})$.
%{These relationships shed further light on how $\textnormal{CC}(p)$
%and its empirical version operate. The form of the $h_{s,j}(\cdot)$'s
%shows that $\textnormal{CC}(\cdot)$ essentially contrasts the reweighted probability mass induced by $\kappa(\cdot)$
%of the interval $(0,p_{s,j}]$ to that of $(p_{s,j},1]$  
%as compared to the $U(0,1)$. 
{Observe that, in view of our nested
partition, increasing $s$ allows for more and more careful checks
of the discrepancies between $F(\cdot)$ and $F_{0}(\cdot)$. More
precisely, we start by considering the deviation at the median of
$F_{0}(\cdot)$, then check the fit at its quartiles and
so on. Also, $\hat{\gamma}_{s,j}$ can be seen as a statistic for testing  $\gamma_{s,j}=0$.}

For $S(n)$ large enough corresponding to $n\geq n_{0}$
say, if $F(\cdot)\neq F_{0}(\cdot)$, there exist $s_{0}\in\{0,1,\ldots,S(n)\}$
and $j_{0}\in\{1,\ldots,d(s_{0})\}$ such that 
\begin{align}
\gamma_{s_{0},j_{0}} & \neq0.\label{eq:non-vanishing CC(p)}
\end{align}

Because we are considering nested partitions, for $n\geq n_{0}$ and
$s\geq s_{0}$, there is a corresponding $j_{0}$ such that (\ref{eq:non-vanishing CC(p)})
remains valid; hence we might as well assume that $n_{0}$, $s_{0},j_{0}$
are the smallest values for which (\ref{eq:non-vanishing CC(p)})
holds.

%% ############    2.4: $\chi^{2}$-type test statistic and selection rule for $d(s)$.  ################

%\subsection{$\chi^{2}$-type test statistic and selection rule for its number of components $d(s)$}

\subsection*{2.4.~~$\chi^{2}$-type test statistic and selection rule for its number of components $d(s)$}

Set
\begin{align}
\mathcal{K}(d(s)) & =n^{1/2}\,\left(\widehat{\textnormal{CC}}(p_{s,1}),\ldots,\widehat{\textnormal{CC}}(p_{s,d(s)})\right)^{\prime}=n^{1/2}\,\left(\hat{\gamma}_{s,1},\ldots, \hat{\gamma}_{s,d(s)} \right)^{\prime}.\label{eq:K-vector of components}
\end{align}
This vector can be seen as  the score vector of an auxiliary parametric model associated with  $\big(h_{1,1}(p),\ldots,h_{s,d(s)}(p)\big)$ modelling an alternative
to $F_0(\cdot)$. 
Consider the  $\chi^{2}$-type test statistic for the GoF
problem of testing $\mathbb{H}_{0}$:
\begin{align}
\mathcal{P}_{d(s)}=\mathcal{K}^{\;\prime}(d(s))\mathcal{K}(d(s)) & =n\,\sum_{j=1}^{d(s)}\left[\widehat{\textnormal{CC}}(p_{s,j})\right]^{2}.\label{eq:test statistic for Case 0}
\end{align}
\textcolor{black}{Note that here and in the sequel, the term ``$\chi^{2}$-type'' refers to the structure of the test statistic, which as in Pearson's standard  $\chi^{2}$ test, is a sum of squares of asymptotically $N(0,1)$ components under  $\mathbb{H}_{0}$, but not to its asymptotic distribution. Indeed, under} 
$\mathbb{H}_{0}$, the null asymptotic distribution of $\mathcal{P}_{d(s)}$
is a sum of weighted $\chi_{1}^{2}$. The covariance matrix \textcolor{black}{$\Lambda_{d(s)}$} of $\mathcal{K}(d(s))$
and the related  score statistic, \textcolor{black}{namely $\mathcal{K}^{\;\prime}(d(s))\big(\Lambda_{d(s)}\big)^{-1}\mathcal{K}(d(s))$,} could be used to obtain a quadratic form with an asymptotic $\chi_{d(s)}^{2}$
distribution. We do not pursue this further because  \textcolor{black}{$\Lambda_{d(s)}$} being
non-diagonal, the components of the score statistic are linear combinations
of the $\widehat{\textnormal{CC}}(p_{s,j})$'s and thus difficult to interpret. Furthermore,
the convenience of a $\chi_{d(s)}^{2}$ reference distribution vanishes
in view of the upcoming enhancements to (\ref{eq:test statistic for Case 0}).

An important question with GoF test statistic (\ref{eq:test statistic for Case 0})
is the proper choice for the number of components $d(s)$ to include.
Here we adapt a data driven selection rule inspired by Ledwina \&
Wy{\l}upek (2012a, 2015) that is defined as follows. First, consider
the auxiliary selection rule with AIC-type penalty
\begin{align}
A(a) & =\min\left\{ d(s)\in\mathbb{D}(n):\mathcal{P}_{d(s)}-a \cdot d(s)\geq\mathcal{P}_{d(t)}-a \cdot d(t),\:d(t)\in\mathbb{D}(n)\right\} .\label{eq:Penalty Case 0}
\end{align}
Now, given $n$ and significance level $\alpha$, find by the Monte
Carlo method a value $a=a(n,\alpha)$ such that, under $\mathbb{H}_{0}$, $\textrm{pr}\big(A(a)=1\big)\geq1-\alpha$.
Such a value exists because  $\textrm{pr}\big(A(a)=1\big)$
is a nondecreasing function of $a\in[0,\infty)$. Then introduce the
auxiliary statistic
\begin{align}
\mathcal{M}_{D(n)} & =\underset{1\leq j\leq D(n)}{\max}\left|n^{1/2}\,\widehat{\textnormal{CC}}(p_{S(n),j})\right|\label{eq:M_D(n) in Case 0}
\end{align}
and denote by $m(n,\alpha)$ the critical value of the $\alpha$-level
test rejecting $\mathbb{H}_{0}$ for large values of $\mathcal{M}_{D(n)}.$
Finally, set
\begin{align}
R(\alpha) & =\begin{cases}
A\big(a(n,\alpha)\big), & \mathcal{M}_{D(n)}\leq m(n,\alpha),\\
A(0), & \mathcal{M}_{D(n)}>m(n,\alpha).
\end{cases}\label{eq:expression for E(alpha)}
\end{align}
With these notations, our data driven GoF $\chi^{2}$-type test statistic
takes the form $\mathcal{P}_{R(\alpha)}$. Its critical values are
obtained via Monte Carlo simulations (see Appendix~B for
some recommendations).

In (\ref{eq:expression for E(alpha)}), test statistic $\mathcal{M}_{D(n)}$
acts like an oracle. When the oracle rejects  $\mathbb{H}_{0}$, i.e. when $\mathcal{M}_{D(n)}  >  m(n,\alpha)$, then
$R(\alpha) = A\big(0\big) =D(n)$ because the  $\mathcal{P}_{d(s)}$  are increasingly ordered. In such cases,  $\mathcal{P}_{R(\alpha)}$ =  $\mathcal{P}_{D(n)}$
and our procedure  seeks confirmation of the oracle's rejection by using the comprehensive test statistic $\text{\ensuremath{\mathcal{P}_{D(n)}}}.$

Now consider the case where $\mathcal{M}_{D(n)}$ accepts the null hypothesis  $\mathbb{H}_{0}$, i.e. $\mathcal{M}_{D(n)}\leq m(n,\alpha)$. This means that 
after examining the large number ($D(n)$) of components in $\mathcal{M}_{D(n)}$, the oracle sees no reason to reject $\mathbb{H}_{0}$ at level $\alpha$.
Then our procedure $\mathcal{P}_{R(\alpha)}$  proceeds to look at a smaller number of components by using the auxiliary selection rule $A(a)$. More
precisely, under $\mathbb{H}_{0}$, it holds that $\textrm{pr}\big(A(a(n,\alpha))=1\big) \geq 1-\alpha$. Hence, the resulting penalty $a(n,\alpha)$ in the selection rule is relatively large. It follows that the (implied) basic selection rule 
$R(\alpha)$ will tend to choose a relatively small dimension $d(s)$. This results in moderately large critical values for  $\mathcal{P}_{R(\alpha)}$, in comparison
with the corresponding critical values for $\mathcal{P}_{D(n)}$. In turn, this leads to more frequent rejections under the alternatives than would produce $\mathcal{P}_{D(n)}$
and thus, results in higher power.

The following proposition is proved in Appendix~D.1.

%\begin{proposition}
%\label{prop: consistency Case 0}Assume that $S(n)\rightarrow\infty$
%and $D(n)$ = $o(n^{2\delta})$ for some $\delta\in(0,1/2)$. Then the
%test rejecting for large values of $\mathcal{P}_{R(\alpha)}$ is consistent
%under any alternative $F(\cdot)\neq F_{0}(\cdot)$.
%\end{proposition}

\begin{customprop}{1}\label{one}
\label{prop: consistency Case 0}Assume that $S(n)\rightarrow\infty$
and $D(n)$ = $o(n^{2\delta})$ for some $\delta\in(0,1/2)$. Then the
test rejecting for large values of $\mathcal{P}_{R(\alpha)}$ is consistent
under any alternative $F(\cdot)\neq F_{0}(\cdot)$.
\end{customprop}

 \textcolor{black}{ In closing this section, note that from the proof of this proposition, it follows that under $\mathbb{H}_{0}$, the first line in (2.9) plays the main role in controlling $R(\alpha)$ while
 under alternatives, the value of $R(\alpha)$ is mainly decided by the second line. }

%% ############    2.5: A simulation experiment  ################ 

%\subsection{A simulation experiment\label{subsec:Simulations case 0}}

\subsection*{2.5.~~A simulation experiment\label{subsec:Simulations case 0}}

\begin{figure}[H]
\begin{centering}
\includegraphics[width = \linewidth]{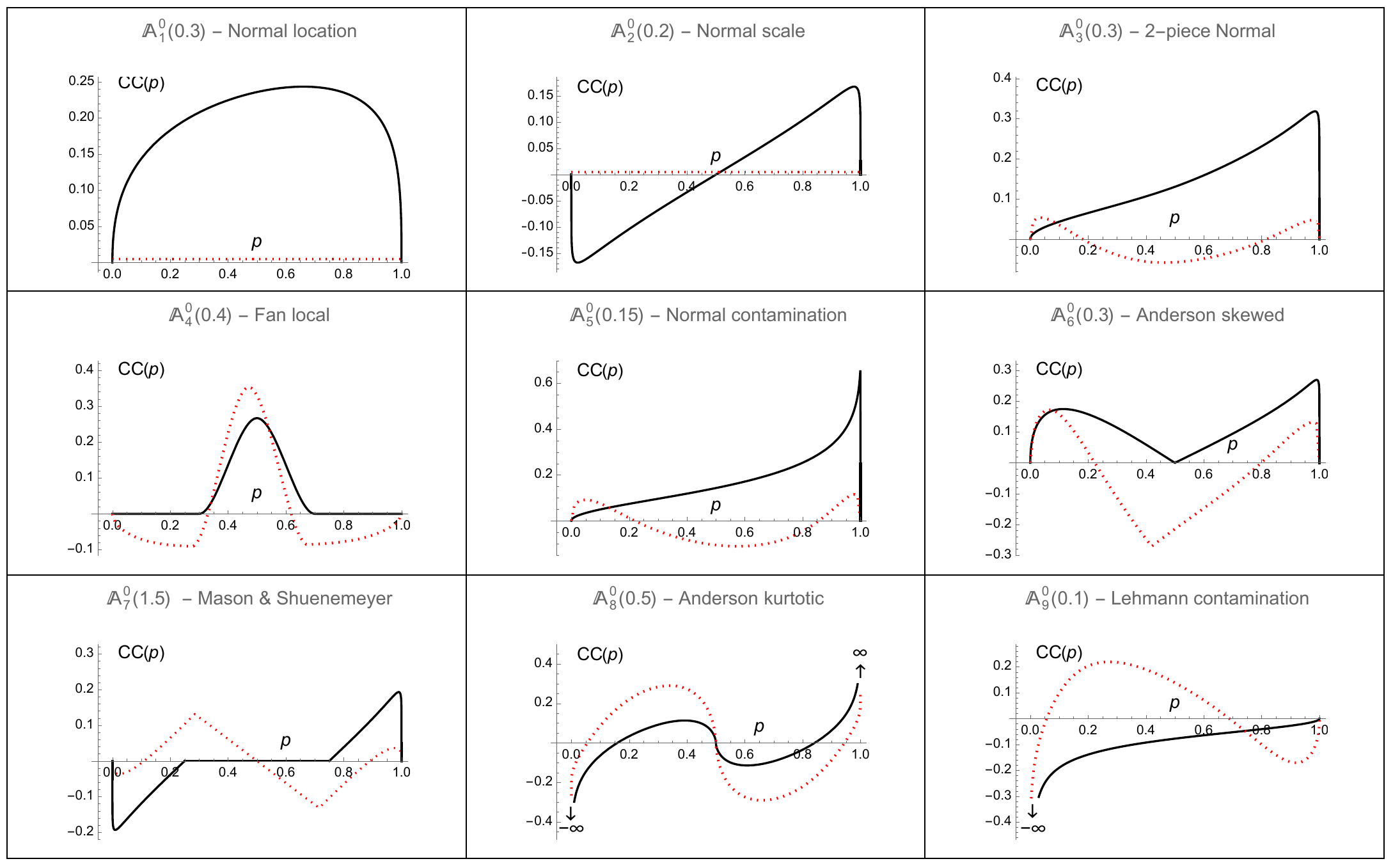}
\caption[wifth=0.5]{\label{fig:CC(curves) Case 0 and 3}$\textnormal{CC}(p)$ (solid black curve) for
the alternative distributions in Table~\ref{tab:Power-at-level 5=000025 Case 0} for testing the simple null hypothesis $\mathbb{H}_{0}: \Phi(x)$.
The red dotted curve represents $\textnormal{CC}(\cdot;\beta(F))$ of (\ref{eq: equation for Gamma_j_0}) corresponding to the null model $\Phi\big((x-\beta_{1})/\beta_{2}\big)$ with $(\beta_{1},\beta_{2}) $ unknown.}
\par\end{centering}
\end{figure}

In order to assess the properties of our test based on $\mathcal{P}_{R(\alpha)}$,
a simulation experiment was performed. The goal was to compare the power of $\mathcal{P}_{R(\alpha)}$ with some of its competitors.
The null hypothesis considered is $F_{0}(x)=\Phi(x)$, the CDF of
the $N(0,1)$ distribution. The alternatives were carefully selected
to cover a fair range of shapes of $\textnormal{CC}(\cdot)$, see Figure~\ref{fig:CC(curves) Case 0 and 3}
and the discussion in Section~4. They are:
\vspace{10pt}

\begin{itemize}
\item $\mathbb{A}_{1}^{0}(\theta)$, a normal location model with CDF $\Phi(x-\theta)$, $\theta\in\mathbb{R}$;
\item $\mathbb{A}_{2}^{0}(\theta)$, a normal scale model with CDF $\Phi(x/(1+\theta))$, $\theta>-1$;
\item $\mathbb{A}_{3}^{0}(\theta)$, the two-piece normal distribution with density  $
C\{I(x<0)\exp(-x^{2}/2) +I(x \geq 0)\exp(-x^{2}/2(1+\theta)^{2})\}$
 with $C=((2\pi)^{1/2}(2+\theta)/2)^{-1}$ and
$\theta>-1$; 
\item $\mathbb{A}_{4}^{0}(\theta)$, a model of Fan with local departure around 0 and density 
$\phi(x)[1+\{4z\theta^{-2}($$\theta-\left|z\right|)\}\,I(\left|z\right|<\theta)]$,
where $z=2\Phi(x)-1$ and $\theta\in[0,1]$;
\item $\mathbb{A}_{5}^{0}(\theta)$, a normal contamination model with CDF $(1-\theta)\Phi(x)+\theta\,\Phi(x-2)$,
$\theta\in[0,1]$;
\item $\mathbb{A}_{6}^{0}(\theta)$, Anderson's skewed distribution with stochastic representation  $ I(Z<0)Z/(1-\theta)+I(Z\geq0)Z(1-\theta)$ 
 where $Z\sim N(0,1)$ and $\theta\in[0,1)$;
\item $\mathbb{A}_{7}^{0}(\theta)$, the Mason \& Schuenemeyer tail alternative with CDF $J(\Phi(x),q, \theta)$ 
 where $\theta > -1$,  $J(x,q, \theta)=\big(q^{\theta / (\theta+1)}  x^{1/(\theta+1)}\big)  I(0\leq x < q)
+x I(q \leq x \leq 1-q) + \big( (1-q^{\theta/(\theta+1)} (1-x)^{1/(\theta+1)}\big)  I(1-q < x \leq 1)$ and here $q=0.25$;
\item  $\mathbb{A}_{8}^{0}(\theta)$,  Anderson's kurtotic distribution generated as $X=Z\cdot\left|Z\right|^{\theta}$,
where $Z\sim N(0,1)$ and $\theta\geq0$;
\item $\mathbb{A}_{9}^{0}(\theta)$, a Lehmann contamination model with CDF $(1-\theta)\Phi(x)+\theta\,(\Phi(x))^{0.175}$,
$\theta\in[0,1]$  with $(\Phi(x))^{\delta}$
being the Lehmann distribution.
\end{itemize}

For more details about these alternatives, see Appendix~C. They all reduce to the $N(0,1)$ when $\theta=0$ and
thus embed the null model. Our choice for $\mathbb{H}_{0}$
offers the convenience of easily defined embedding families of alternatives
and allows some comparisons with the simulations in Appendix~C.
No loss of generality ensues from this choice as the probability integral
transformation translates the GoF problem for any continuous $F_{0}(\cdot)$
into a null $U(0,$1) distribution, for which many good tests have
been derived. In particular, we have considered the following two
competitors: the Anderson-Darling statistic (AD) and the Berk \& Jones (1979) statistic (BJ).

Taking $\alpha=0.05$ and $n=100$, we investigated $\mathcal{P}_{R(\alpha)}$
with $S(n)=6$ and computed $m(n,\alpha)=3.30$ and $a(n,\alpha)=3.31$ (see Appendix~B for some practical considerations
regarding these choices and values).
The power functions for the nine alternatives were simulated in their
$\theta$ range for each of AD, BJ and $\mathcal{P}_{R(\alpha)}$.
The 5\% critical values were obtained from 100~000 replications under
the null distribution, while the powers were computed from 10~000 Monte
Carlo runs. We extracted from each power curve a representative value
of $\theta$ which provided intermediate powers, e.g. not too close
to 0.05 and 1.0 and where the powers of the various tests could be distinguished.
As a result of these choices, the powers presented below offer a broad
view of the comparative behaviour of the tests over a fair range
of situations. 

The results are reported in Table~\ref{tab:Power-at-level 5=000025 Case 0}. None of the tests  dominates and $\mathcal{P}_{R(\alpha)}$
emerges as very competitive. In most cases, AD is  less powerful and simulations results in  \'Cmiel, Inglot \& Ledwina  (2020) show
that for larger sample sizes and heavy tails, the power differences between AD and $\mathcal{M}_{D(n)}$ can be even more pronounced. 
Thus powerwise, we conclude that our
test could be included among the best solutions for this problem.

We close this section by noting that when any of
AD, BJ rejects $\mathbb{H}_{0}$, the user is left with little clues
as to what aspects of the null model must be corrected. In contrast,
when the competitive $\mathcal{P}_{R(\alpha)}$ rejects, a B-plot
as in Panel 2) of Figure~\ref{fig:miling baby data set} can be produced
to help the user see where the, now statistically established, discrepancies are
located. The B-plot could be further supplemented with acceptance regions  to separate in a reasonable way
the large components  from those more consonant with a local agreement
to the model. This will be discussed in Section~3.3 and illustrated in Section ~3.5 (see also Appendix~A). Hence the pair (B-plot, $\mathcal{P}_{R(\alpha)}$) can be a useful tool for statistical modelling.

We now
consider the case where the parameter $\beta$ in model
$F_{0}(\cdot\,;\,\beta)$ is unknown.

\begin{table}[H]
\begin{centering}
\begin{tabular}{ccccc}
Alternative  & AD & BJ & $\mathcal{M}_{127}$ & $\mathcal{P}_{R(\alpha)}$ \\[3pt]
\hline 
$\mathbb{A}_{1}^{0}(0.3)$ & 83 & 69 & 65 & 77 \\
$\mathbb{A}_{2}^{0}(0.2)$ & 35 & 51 & 59 & 58 \\
$\mathbb{A}_{3}^{0}(0.3)$  & 67 & 73 & 78 & 79\\
$\mathbb{A}_{4}^{0}(0.4)$  & 27 &  47 &  39 & 39\\
$\mathbb{A}_{5}^{0}(0.15)$ & 83 & 92 & 93 & 94\\
$\mathbb{A}_{6}^{0}(0.3)$ & 49 & 65 & 63 & 64\\
$\mathbb{A}_{7}^{0}(1.5)$ & 24 & 54 & 63 & 59\\
$\mathbb{A}_{8}^{0}(0.5)$ & 47 & 88 & 80 & 80\\
$\mathbb{A}_{9}^{0}(0.1)$ & 53 & 90 & 79 & 75 \\[5pt]
\end{tabular}
\par\end{centering}
\caption{\label{tab:Power-at-level 5=000025 Case 0}Powers ($n=100$, $\alpha=0.05$) of the Anderson-Darling
 (AD), the Berk \& Jones (1979)  (BJ), the oracle $\mathcal{M}_{D(n)}$ with $D(n)=127$ and our $\mathcal{P}_{R(\alpha)}$ tests
for $\mathbb{H}_{0}:F(x)=\Phi(x)$ against a set of balanced alternatives.}
 
\end{table}

%% ############   Section 3.0   Composite null hypothesis  ################

\section{Composite null hypothesis\label{sec:Section 3 : Composite-null-hypothesis}}

In the case of a composite null hypothesis, we proceed in a reversed
order than under $\mathbb{H}_0$. We start with a counterpart to the empirical Fourier coefficients
(FC) associated with an adjusted variant of $\hat\gamma_{s,j}$ to motivate
our definition of an analogue to $\textnormal{CC} (\cdot)$ in the
present setting. Next, we define the related $\chi^{2}$-type test
and its data driven version. We focus on the important case of location-scale
families and illustrate our approach by testing GoF to the Gaussian
distribution. A simulation experiment, reported in Appendix~C, confirms that our  test
is competitive in a broad spectrum of situations. Finally, we produce
the B-plots of four real data sets (more are worked out in Appendix~A) to show their usefulness
in obtaining  insights about the aspects of the model that
could have caused rejection or led to an erroneous conclusion because of the application of an inadequate GoF test.

%% ############   Section 3.1   Fourier coefficients and the $\chi^{2}$-type test statistic  ################

%\subsection{Fourier coefficients and empirical CC}
\subsection*{3.1.~~Fourier coefficient, empirical CC and B-plot}

Let $X_{1},\ldots,X_{n}$ be i.i.d. observations from a continuous CDF
$F(\cdot)$ and consider the family of null models $\text{\ensuremath{F_{0}}(\ensuremath{\cdot}\,
;\,\ensuremath{\beta})}$
where the Euclidean parameter $\beta$ $\in\mathcal{B}$ is unknown. The null
hypothesis of interest is composite and takes the form $\mathbb{H}:F(\cdot)=F_{0}(\cdot\,;\,\beta) \;\textrm{for some unknown }\beta\in\mathcal{B}$.

As in Section~2.3, the
$(s,j)$-th empirical FC of the comparison density associated with $F(\cdot)$ and $F_{0}(\cdot\,;\,\beta)$ can be defined as
\begin{align*}
\hat{\gamma}_{s,j} (\beta)= \hat{\gamma}_{s,j}(p_{s,j};\,\beta)  &=\frac{1}{n}\sum_{i=1}^{n}h_{s,j}(F_{0}(X_{i};\beta)).
\end{align*}
Plugging into this expression the value $\tilde{\beta}$ of
an estimator of $\beta$, elementary calculations
yield
\begin{align}
\hat{\gamma}_{s,j} & (\tilde{\beta})=\frac{p_{s,j}-\bar{F}_{n}(p_{s,j};\tilde{\beta})}{(p_{s,j}(1-p_{s,j}))^{1/2}}, &\quad  \bar{F}_{n}(p;\tilde{\beta}) & =n^{-1} \sum_{i=1}^{n}I(F_{0}(X_{i};\tilde{\beta})\leq p). \label{eq:empirical FC Case 3}
\end{align}
It is tempting to define the empirical CC as  (\ref{eq:empirical FC Case 3}). However, the use of $\tilde{\beta}$
must be taken into account. Consider the empirical process $\bar{e}_{n}(p;\tilde{\beta}) =n^{1/2}(\bar{F}_{n}(p;\tilde{\beta})-p),\;p\in[0,1]$.
Durbin (1973) has studied this process and shown that under mild smoothness assumptions on $F_{0}(\cdot\,;\,\cdot)$ \textcolor{black}{(see his assumption A.2)} and if $\tilde\beta$ satisfies \textcolor{black}{   $n^{1/2}( \tilde{\beta}-\beta)=n^{-1/2} \sum_{i=1}^n B(X_i,\beta)+o_p(1)$, where $B(X,\beta)$ has mean 0 and a finite covariance matrix under $\mathbb{H}$  (a variant of his assumption A.1 adapted to our context)}, this process converges under
$\mathbb{H}$ to some Gaussian process on $[0,1]$ with 0 mean function.
The covariance function $\rho(\cdot,\cdot)$ of the limiting process
is also given in Durbin (1973). We assume throughout that Durbin's assumptions hold true. Hence, putting $\sigma_{s,j}=(\rho(p_{s,j},p_{s,j}))^{1/2}$,
we get that $\bar{e}_{n}(p_{s,j};\tilde{\beta})/\sigma_{s,j}$ is
asymptotically $N(0,1)$ under $\mathbb{H}$. We also refer  to Neuhaus (1979) for a thorough discussion of Durbin's (1973) paper.

The above leads to defining
the empirical CC in the present context as 
\begin{align*}
\widehat{\textnormal{CC}}(p;\tilde{\beta}) & =\frac{p-\bar{F}_{n}(p;\,\tilde{\beta})}{\rho^{1/2}(p,p)}.
\end{align*}
The graph of $\widehat{\textnormal{CC}}(\cdot;\,\tilde{\beta})$ evaluated on the points $p_{s,j}$'s of a grid defines the B-plot
when the nuisance parameter $\beta$ is present.

%% ############   Section 3.2   Alternatives and \emph{CC} in the case of location-scale models  ################

%\subsection{\label{subsec:Alternatives-}Alternatives and CC in the case of location-scale models}{\label{subsec:Alternatives-}
\subsection*{3.2.~~CC and alternative in the case of location-scale models}{\label{subsec:Alternatives-}

Let $F(\cdot)$ be the true CDF of the data and suppose the null model
is location-scale with $\beta$ estimated
by $\ensuremath{\tilde{\beta}}$, a $n^{1/2}$-consistent estimator
under $F_{0}(\cdot\,;\cdot)$. Suppose that under $F(\cdot)$, $\ensuremath{\tilde{\beta}}$
$\ensuremath{\rightarrow}\beta(F)=(\beta_{1}(F),\beta_{2}(F))^{\prime}$
where the convergence is in probability with respect to $F(\cdot)$.
The population version of $\widehat{\textnormal{CC}}(p;\tilde{\beta})$  is
\begin{align}
\textnormal{CC}(p;\beta(F))= & \frac{p-F\Bigl(\beta_{2}(F)\cdot F_{0}^{-1}(p)+\beta_{1}(F)\Bigr)}{\rho^{1/2}(p,p)}. \label{eq: equation for Gamma_j_0}
\end{align}
We call $F(\cdot)$ an alternative when $F(x)\neq  F_{0}\big(({x-\beta_{1}(F)})/{\beta_{2}(F)}\big)$
for some $x\in\mathbb{R}.$ 
By continuity of $F(\cdot)$ and the fact
that the partition of Section~2.2
is dense, there exists $\ensuremath{n_{0},\;s_{0}\in\{0,\ldots,S(n_{0})\}\;}$
and $\ensuremath{j_{0}\in\{1,\ldots,d(s_{0})\}}$, such that $\textnormal{CC}(p_{s_{0},j_{0}};\beta(F))\neq0$. Without loss of generality we may, 
as in Section~2.3,
assume that $\ensuremath{n_{0},s_{0}},j_{0}$ are the smallest such indices. Thus for sufficiently
large $\ensuremath{n}$ and under $F(\cdot)$, there will be at least one non-zero component
among the $\textnormal{CC}(p_{s,j};\beta(F))$'s.

%% ############   Section 3.3 Using the pair (\mathcal{P}_{\tilde{R}(\alpha)}(\widetilde{\beta})$, bar disply) he composite case  ################

%\subsection{The bar plot (B plot) in the composite and simple null cases\label{subsec:Simulations Case 3}}
\subsection*{3.3.~~Acceptance region for subset of bars\label{subsec:Simulations Case 3}}

Similarly to the case of Section~2.1 where $\beta(F)$ is known,  $\textnormal{CC}(p; \beta(F))$ inherits its interpretation from its local sign and possible zeroes. Now $F_0(\cdot;\,\beta(F))$ is the reference CDF. Therefore, it is important to discriminate  those bars that seem compatible with $\mathbb{H}$ from more surprisingly large ones,  positive or negative. This suggests, as a first step, supplementing the B-plot with one-sided $(1-\alpha)$-th acceptance intervals for the height of an individual bar expected under the null model. Here, this is done by drawing on the B-plot horizontal lines at $\pm 1.645$, if one considers  $\alpha = 0.05$. We use asymptotic critical levels because they approximate well the finite sample distribution of single bars; see Section~3.5.1.
 \textcolor{black}{ Note that to ensure the statistical rigor in a classical sense of such acceptance regions, the position of the bar of interest and its direction (i.e. lower or upper  region) should be determined by external considerations or previous knowledge, not from  looking at the B-plot arising from a given sample. }
 %However, if such outside information is not %available, the acceptance region of a bar can still be used in an informal way to suggest some insight about the deficiencies of the null model.}

\textcolor{black}{Insight deriving from a single bar may be rather limited and, in view of the interpretation of B-plots as explained below equation (\ref{eq:CC(p) curve)}), instead of a single bar, one may be interested in subsets of adjacent bars and there simultaneous acceptance regions. Of course the same caveat as above applies about their uses in practice.} Now for such subsets,  recall  from Durbin's (1973) result that the joint asymptotic distribution of a subset of bars is multivariate normal with means 0, unit variances and associated covariance function. Thus the correlation between the bars must be taken into account.  When  a subset of bars, say $n^{1/2}\,\widehat{\textnormal{CC}}(p_{S(n),j};\tilde{\beta})$ for all $ j\in  \mathcal{J}$, \textcolor{black}{are expected to be} jointly positive, one approach to flag their  significance  is to compute a  one-sided  simultaneous $1-\alpha$ level  acceptance region. This consists in computing by  Monte Carlo  and under $\mathbb{H}$ an approximation to $u(n, \alpha;\mathcal{J})$ in
\begin{align}
\textrm{pr}\Bigl(\textrm{max}_{j\in  \mathcal{J}} ~n^{1/2} \,\widehat{\textnormal{CC}}(p_{S(n),j};\tilde{\beta}) \leq u(n, \alpha; \mathcal{J})\Big) \geq 1-\alpha. \label{eq:p-value case 3}
\end{align}
If the bars under consideration should be negative,  (\ref{eq:p-value case 3}) must be adapted to obtain the lower bound $\ell(n, \alpha;  \mathcal{J})$ via
\begin{align*}
\textrm{pr}\Bigl(\textrm{min}_{j\in  \mathcal{J}} ~n^{1/2} \,\widehat{\textnormal{CC}}(p_{S(n),j};\tilde{\beta}) \geq \ell(n, \alpha;   \mathcal{J})\Big) \geq 1-\alpha.
\end{align*}
If desired, a two-sided simultaneous $1-\alpha$ level acceptance region for all $j \in \mathcal{J}$ can be defined as follows. With $u(n,\alpha;\mathcal{J})$ and $\ell(n,\alpha;\mathcal{J})$ as defined above, it holds that 
$$
\text{pr}\Bigl( \min_{j \in \mathcal{J}} \sqrt n \widehat{\text{CC}}(p_{S(n),j}) \geq \ell(n,\alpha;\mathcal{J}),\; \max_{j \in \mathcal{J}} \sqrt n \widehat{\text{CC}}(p_{S(n),j}) \leq u(n,\alpha;\mathcal{J}) \Bigr) \geq 1-2\alpha.
$$
Hence $[\ell(n,\alpha/2;\mathcal{J}),u(n,\alpha/2,\mathcal{J})]$ forms a two-sided $1-\alpha$ level acceptance regions for bars in $\mathcal{J}$.

Such computations are easily done because with a location-scale null model and a  $\tilde{\beta}$ invariant to location-scale transformations, the distribution of $\widehat{\textnormal{CC}}(p_{S(n),j};\tilde{\beta})$ does not depend on $\beta$, 
so sampling can be made from e.g. $F_0(\cdot\; ; (0, 1))$.  When $\beta$ is known, a variant of $u(n,\alpha,  \mathcal{J})$ can be computed by approximating 
under the null hypothesis  expression (\ref{eq:p-value case 3}) using  $\widehat{\textnormal{CC}}(p_{S(n),j})$, and similarly for  $\ell(n,\alpha,  \mathcal{J})$.   
When at least one of the bars related to an element of $j\in  \mathcal{J}$ is above the computed $u(n, \alpha;  \mathcal{J} )$ or below  $\ell(n, \alpha;  \mathcal{J})$, 
the heuristic suggests that the data seems incompatible with the null model in the related region. 
We  represent these simultaneous acceptance regions by shaded grey stripes on the B-plot, see Section~3.5 or Appendix~A 
 for illustrations \textcolor{black}{and Appendix~C for some information about the required  computational effort}.
\textcolor{black} {Also, in these computations the question of choosing appropriately the simultaneous level $\alpha$ of such acceptance regions arises.  See  Section~3.5  for some discussion about this point.}
 
In the examples below,  we have considered data sets that have  been previously studied and discussed. This previous knowledge justifies the application of one-sided simultaneous acceptance regions in order to check if our conclusions support existing ones, if our methods provide additional insights into the structure of these data, if some explanations can be offered as to why some known tests have failed to work, etc. Such settings implies the use of some external information to define the sets of bars under consideration. But in many applications, only a few prescribed regions of population quantiles will be of interest, \textcolor{black}{e.g. the extreme deciles for  tail regions or the central tercile for the centre of the distribution.}

When such external information is not be available,  one needs some objective way to define the sets of bars to further study in order to get insights. As an example of what can be done, and based  on statistical practice in the two-sample setting, Ledwina and Zagda\'nski (2024) have proposed, in a related study, to split the range [0,1] of population quantiles into 10 equal length intervals and to consider subsets of bars falling into these intervals. They illustrate and discuss the approach for data on income and cholesterol levels. A similar approach can be adapted to the present context of goodness of fit testing. See Appendix A.3 for an illustration. \\
%% ############   Section 3.4 Selection rule for $d(s)$ in the case of  testing Gaussianity  ################

%\subsection{$\chi^2$-type test statistic and a selection rule for $d(s)$ for testing Gaussianity}
\subsection*{3.4.~~$\chi^2$-type test statistic and a selection rule for $d(s)$ for testing Gaussianity}

Consider as in (\ref{eq:K-vector of components}), $\mathcal{K}(\tilde{\beta},d(s))=n^{1/2}\left(\widehat{\textnormal{CC}}(p_{s,1};\tilde{\beta}),\ldots,\widehat{\textnormal{CC}}(p_{s,d(s)};\tilde{\beta})\right)^{\prime}$.
Inspired by (\ref{eq:test statistic for Case 0}), a GoF test statistic
for $\mathbb{H}$ is
\begin{align}
\mathcal{P}_{d(s)}(\tilde{\beta}) & =\mathcal{K}^{\;\prime}(\tilde{\beta},d(s))\mathcal{K}(\tilde{\beta},d(s))=n\sum_{j=1}^{d(s)}\left[\widehat{\textnormal{CC}}(p_{s,j};\tilde{\beta})\right]^{2}.\label{eq: Test statistic Case 3 general siruation}
\end{align}

\begin{customremark}{1}
When $F_{0}(\cdot\,;\,\beta)$ is a location-scale family, i.e. $F_{0}(x;\,\beta)=F_{0}\big(({x-\beta_{1}})/{\beta_{2}}\big)$
with $\beta=(\beta_{1},\beta_{2})^{\prime}\in\mathcal{B}\subseteq\mathbb{R}\times\mathbb{R}_{+}$,
we have $\bar{F}_{n}(p;\,\tilde{\beta})=\hat{F}_{n}(\tilde{\beta}_{2}F_{0}^{-1}(p)+\tilde{\beta}_{1}),$
where $\hat{F}_{n}(\cdot)$ is the ordinary empirical CDF of the sample. Under Durbin's assumptions,
the process $\bar{e}_{n}(p;\tilde{\beta})$ has a limiting distribution
that does not depend on $\beta$. Test statistic (\ref{eq: Test statistic Case 3 general siruation})
becomes 
\begin{alignat}{1}
\mathcal{P}_{d(s)}(\tilde{\beta})=n\sum_{j=1}^{d(s)}\left[\frac{p_{s,j}-\hat{F}_{n}(\tilde{\beta}_{2}F_{0}^{-1}(p_{s,j})+\tilde{\beta}_{1})}{\sigma_{s,j}}\right]^{2}.\label{eq:test statisitc- location scale case}
\end{alignat}
\end{customremark}

To reduce technicalities, consider from now on the important sub-case where $F_{0}(x;\,\beta)=F_{0}\big(({x-\beta_{1}})/{\beta_{2}}\big)$
is the Gaussian CDF with unknown expectation $\beta_{1}$ and variance
$\beta_{2}^{2}$ , i.e. $F_{0}(\cdot)=\Phi(\cdot)$ with density $f_{0}(\cdot)=\varphi(\cdot)$.
To estimate $\beta_{1}$ and $\beta_{2}^{2}$, we consider the maximum likelihood estimators (MLE), i.e.  $\tilde{\beta}_{1}=\bar{X}$
and $\tilde{\beta}_{2}^{2}=S^{2}=n^{-1}\sum_{i=1}^{n}(X_{i}-\bar{X})^{2}$. This model and these estimates satisfy all assumptions in Durbin's theorem. Hence
we have $\rho(t,v)  =\min\{t,v\}-tv-\rho_{1}(t)\rho_{1}(v)-\rho_{2}(t)\rho_{2}(v), ~t,v\in[0,1],$
where $\rho_{1}(t)=\varphi\bigl(\Phi^{-1}(t))$ and $\;\rho_{2}(t)=2^{-1/2}~\varphi\bigl(\Phi^{-1}(t))\Phi^{-1}(t)$.
In this case, the
quantity $\sigma_{s,j}=(p_{s,j}(1-p_{s,j})-\rho_{1}^{2}(p_{s,j})-\rho_{2}^{2}(p_{s,j}))^{1/2}$
is smaller than the value $(p_{s,j}(1-p_{s,j}))^{1/2}$ adequate for
$\mathbb{H}_{0}$. To be specific, $\rho(t,t)$  \textcolor{black}{ is symmetric about $1/2$,  $\displaystyle \cap$ shaped but} much smaller
than $t(1-t)$ in $(0,1),$ while $\ensuremath{\lim_{t\to0+}\rho(t,t)/\big(t(1-t)\big)=\lim_{t\to1-}\rho(t,t)/\big(t(1-t)\big)=1}$.

To define a selection rule for $\ensuremath{d(s)}$ in (\ref{eq:test statisitc- location scale case}),
we proceed similarly as in Section~2.4.
First, we seek an oracle test whose task is to provide
some reliable preliminary information on the situation. The results of Section~\ref{sec: 2 Case 0} suggest considering the following version of the oracle test (\ref{eq:M_D(n) in Case 0}), namely 

\begin{align}
\mathcal{M}_{D(n)}(\tilde{\beta}) & =\underset{1\leq j\leq D(n)}{\max}\left| n^{1/2}\,\widehat{\textnormal{CC}}(p_{S(n),j};\tilde{\beta})\right|\label{eq:M_D(n) in Case 3}.
\end{align}
However, it turns out (see Table C.2 in Appendix~C), that this test is powerwise inferior to some recommended test procedures in the  composite context. 
The use of selection rules based on this oracle leads to data driven tests which are noticeably more powerful than $\mathcal{M}_{D(n)}(\tilde{\beta})$, but still not competitive with the best existing solutions.  This is a display of the difficulties encountered in moving from a simple to a composite null hypothesis.

Instead, we base our selection rule on the oracle test 
\begin{alignat*}{1}
\mathcal{R}_{n} & =1-\frac{\hat{\sigma}_{n}^{2}}{S^{2}},\quad \hat{\sigma}_{n}=\intop_{0}^{1}\hat{F}_{n}^{-1}(t)\,\Phi^{-1}(t)\,dt,
\end{alignat*}
where large observed values of $\mathcal{T}_{n}=n\mathcal{\,R}_{n}$
are significant. This test statistic, which we refer to as BCMR, has been introduced in del Barrio, Cuesta-Albertos, Matran \& Rodriguez (1999) 
and further studied in Cs\"{o}rg\H{o} (2003), among others. The surrounding theory regarding this test allows its adaptation to some other composite
null hypotheses than the Gaussian. Hence, the solution below can serve as a template  in a variety of important cases.

Introduce
\begin{alignat*}{2}
A(a;\tilde{\beta}) & =\min\left\{ d(s)\in\mathbb{D}(n):\mathcal{P}_{d(s)}(\tilde{\beta})-a\cdot d(s)\geq\mathcal{P}_{d(t)}(\tilde{\beta})-a\cdot d(t),\:d(t)\in\mathbb{D}(n)\right\}.
\end{alignat*}
Now, given $\ensuremath{n}$ and $\ensuremath{\alpha}$, find 
by the Monte Carlo method a value $\ensuremath{a=a(n,\alpha;\tilde{\beta})}$
such that, under $\ensuremath{\mathbb{H}}$,  $\ensuremath{\textrm{pr}(A(a(n,\alpha;\tilde{\beta}))=1)\geq1-\alpha}$.
Finally, let $t(n,\alpha)$  be the  $\alpha$-level critical value of ${\mathcal T}_n$ and  set 
\begin{align*}
\tilde{Q}(\alpha) & =\begin{cases}
A\big(a(n,\alpha;\tilde{\beta}\big); \tilde{\beta}), & \mathcal{T}_{n}\leq t(n,\alpha),\\
A(1.5;\tilde{\beta}), &  \mathcal{T}_{n}>t(n,\alpha).
\end{cases}
\end{align*}
Notice that the penalty in the case $\mathcal{T}_{n}> t(n,\alpha)$ differs from that in (\ref{eq:expression for E(alpha)}). The reason for this is explained in Appendix~B.  $\tilde{Q}(\alpha)$ can be seen as an adaptation of the selection rule $A$ in Ledwina \& Wy{\l}upek (2015) introduced in the context of data driven test associated with transformed Hermite polynomials.

With these notations, the data driven GoF $\chi^{2}$-type test statistic
for the null hypothesis $\mathbb{H}$ takes the form $\mathcal{P}_{\tilde{Q}(\alpha)}(\tilde{\beta})$.
Some of its critical values $\tilde{c}(n,\alpha)$ are listed in  Appendix~B
 and others can be obtained via linear interpolation or Monte Carlo simulations. The following proposition and remark are proved 
 in Appendix~D.2 and Appendix~D.3 respectively.

\begin{customprop}{2}\label{two}
\label{prop: consistency Case 3-1}Let $F(\cdot)$ be an alternative
in the sense of Section 3.2 to the Gaussian
null model. Assume that the fourth moment of $F(\cdot)$ exists and
is finite. Moreover, assume that  $F(\cdot)$  possesses a bounded density $f(\cdot)$ with respect to the Lebesgue measure on $\mathbb R$.
Further assume that $\ensuremath{\tilde{\beta}}$ is the MLE for $\ensuremath{\beta}$.
Finally, let $S(n)\rightarrow \infty$ and  $D(n)= o(n^{1/2})$ as $n \rightarrow \infty$. Then, the test rejecting for large values
of $\mathcal{P}_{\tilde{Q}(\alpha)}(\tilde{\beta})$ is consistent under $F(\cdot)$.
%\end{proposition}
\end{customprop}

\begin{customremark}{2}\label{remark on conv}
Under the assumptions on $F(\cdot)$ and  $\tilde\beta$ in Proposition \ref{two},
it holds that, as $n \rightarrow \infty$,  $\sup_{\epsilon \leq p \leq 1-\epsilon} \bigl|\widehat{\textnormal{CC}}(p;\tilde\beta)-\textnormal{CC}(p;\beta(F)) \bigr| \rightarrow 0$ in probability for any $\epsilon \in (0,1)$. 
\end{customremark}

In order to assess the properties of the test based on $\mathcal{P}_{\tilde{Q}(\alpha)}(\widetilde{\beta})$,
a simulation experiment was performed. The structure of the experiment
mimics closely that in Section~2.5. The
null hypothesis is $\mathbb{H}:F(x)=\Phi\big((x-\beta_{1})/{\beta_{2}}\big)$
where $(\beta_{1},\beta_{2}^2)$ are estimated by MLE. For this problem,
several solutions exist (see Arnastauskait\'e, Ruzgas \& Braz\'enas, 2021), notably the Anderson-Darling (AD), the Shapiro-Wilks (SW)
and the  BCMR tests.  

Our main interest  is to see how our approach compares with these.
The details and results of the simulation appear in Appendix~C, which pertains to a set of carefully selected alternatives 
according to the form of their $\textnormal{CC}(\cdot~;\beta(F))$ partly inspired by those in Section~2.5.
It emerges from this experiment that, as an oracle, $\mathcal{M}_{D(n)}(\tilde{\beta})$ generally does poorly. Otherwise, and similarly to the context 
of Section~\ref{sec: 2 Case 0},  none of the other tests dominates
and $\mathcal{P}_{\tilde{Q}(\alpha)}(\widetilde{\beta})$ turns out to be 
a good competitor, being powerwise on par with the oracle test it is based upon. Recall that a
main advantage of our approach is the possibility of deriving information from the B-plot about where the null model could be at fault and an overall measure of fit based on this plot.
Note however that regarding the tests of $\mathbb{H}_0$ and $\mathbb{H}$, this common approach allows to exhibit some essential differences between the two problems. See
the discussion in Appendix~C.

%% ############   An example  ################

\subsection*{3.5.~~Real data examples \label{subsec:Real-data-examples}}

The results of the previous sections are now applied to some real data
sets to show how useful insights can be derived from the components $n^{1/2}\,\widehat{\textnormal{CC}}(\cdot\,;\,\tilde{\beta})$
when using the methods of the paper. More examples are worked out in Appendix~A. \textcolor{black}{Programs in the Mathematica language
(Wolfram Research, Inc., Mathematica, Version 12.1, Champaign, IL, 2020)  to compute the
new data-driven test statistic, the B-plots and the acceptance regions can be found in the GITHUB repository gilles-ducharme/GoF.Validation.}

\subsubsection*{3.5.1~The wave records data }

We consider a data set in Bickel and Doksum (1977, p. 384, Table 9.6.3) 
measuring the time spent above a high level of $n=66$ wave records in the San Francisco bay. 
Their analysis does not reject the null Gaussian hypothesis at level 10\%.  The data were
 later considered  by Rosenkranz (2000) whose simultaneous $90\%$ confidence band approach indicates some
inconsistency with the postulated Gaussian model in the left tail of the distribution. They were reexamined
by Aldor-Noiman et al. (2013) using simultaneous confidence bands
about the QQ plot. Their approach detects, at the 5\% level, a significant
departure from normality in the right tail, while another approach, based on Kolmogorov-Smirnov bands, 
nearly rejects, at the same level, relying solely on
 points at the centre of the data.

For this data set, we  have $\tilde\beta_1=\bar{X}=3.79$ and $\tilde\beta_2= S = 2.39$.
We apply the test based on $\mathcal{P}_{\tilde{Q}(\alpha)}(\tilde{\beta})$ with $\alpha=0.05$,  
$S(n)=4$, $t(n,\alpha) = 2.60$ and compute $a(n,\alpha;\tilde\beta) = 3.18$. 
The observed value of  $\mathcal{T}_{n}$ is 4.52, leading to $\tilde{Q}(\alpha) $ = 31 and  $\mathcal{P}_{\tilde{Q}(\alpha)}(\tilde{\beta})$ = 92.40 to be compared to 
the $5\%$ critical level of 10.46 obtained from 100 000 Monte Carlo replications. Thus we reject the null hypothesis of Gaussianity at the $5\%$ level.
The tests AD, SW and BCMR also reject at the $5\%$ level with  $p-$values of 0.004 for AD, 0.002 for SW  and 0.002 for BCMR.
Figure~\ref{fig:Empirical-FC for the B=000026 D wave data set}
shows the B-plot of  $n^{1/2}\,\widehat{\textnormal{CC}}(p_{4,j};\tilde{\beta})$
with $j\in\{1,\ldots,31\}$ along with the $\pm1.645$ lines delimiting the one-sided upper and lower $0.05-$level asymptotic individual acceptance regions (dashed horizontal  lines in Figure~\ref{fig:Empirical-FC for the B=000026 D wave data set}).

The value of $n^{1/2}\,\widehat{\textnormal{CC}}(p_{4,31};\tilde{\beta})$
above the dashed line is consonant with the finding of Aldor-Noiman
et al. (2013) of a fatter right tail than a Gaussian distribution.
Assuming that the meaning of their  term ``right tail''  relates to quantiles such that $p \geq 29/32 \approx 0.9$,  we can substantiate this
by computing from (\ref{eq:p-value case 3}) $u(66,0.05; \{ 29,30,31 \}) = 2.21$. From Figure~\ref{fig:Empirical-FC for the B=000026 D wave data set}, because the bar at $p=31/32$ is above 2.21, this further supports the claim of Aldor-Noiman et al. (2013). 
If we define similarly the left tail as below the  first decile, i.e. $p \leq 3/32$, we get $u(66,0.05; \{ 1,2,3 \}) = 1.96$. Again from Figure~\ref{fig:Empirical-FC for the B=000026 D wave data set}, all three bars in $\{ 1,2,3 \}$ are above this value,  in agreement with  the finding in Rosenkranz (2000)  regarding a thinner than Gaussian left tail.

Finally, Aldor-Noiman et al. (2013) do not define the meaning of ``centre of the data'' but, in view of the above findings of a thinner left and fatter right tail, suggesting some asymmetry to the right, we have decided to consider a slightly  shifted centre ranging from  $0.45 \approx 15/32 \leq p \leq 21/32 \approx 0.65$. This yields $\ell(66, 0.05; \{ 15,\ldots,21 \})=-2.07$. The fact that three bars are below this value  is consonant with the claim that, in this area, observations tend to be stochastically smaller than expected under Gaussianity.  

\textcolor{black}{Note that in computing the above acceptance regions,  we have used the same level $\alpha$ ( = $5\%$  here) as in the global test procedure. The main goal at this stage of the analysis is to derive heuristic, but principled, informations as to whether the departures are located where we believe they could be. Therefore, it seems reasonable to use for each subset of bars the initial level  $\alpha$. This is in accordance with similar work in extracting diagnostic information following the rejection of a null model, as in Thas (2010, Sections 4.2.1.2 and 4.2.1.3). Thus in the sequel, all acceptance regions have been computed at level 5\% for each subset of bars considered. }

 \begin{figure}[H]
\begin{centering}
\includegraphics[width =0.75\linewidth]{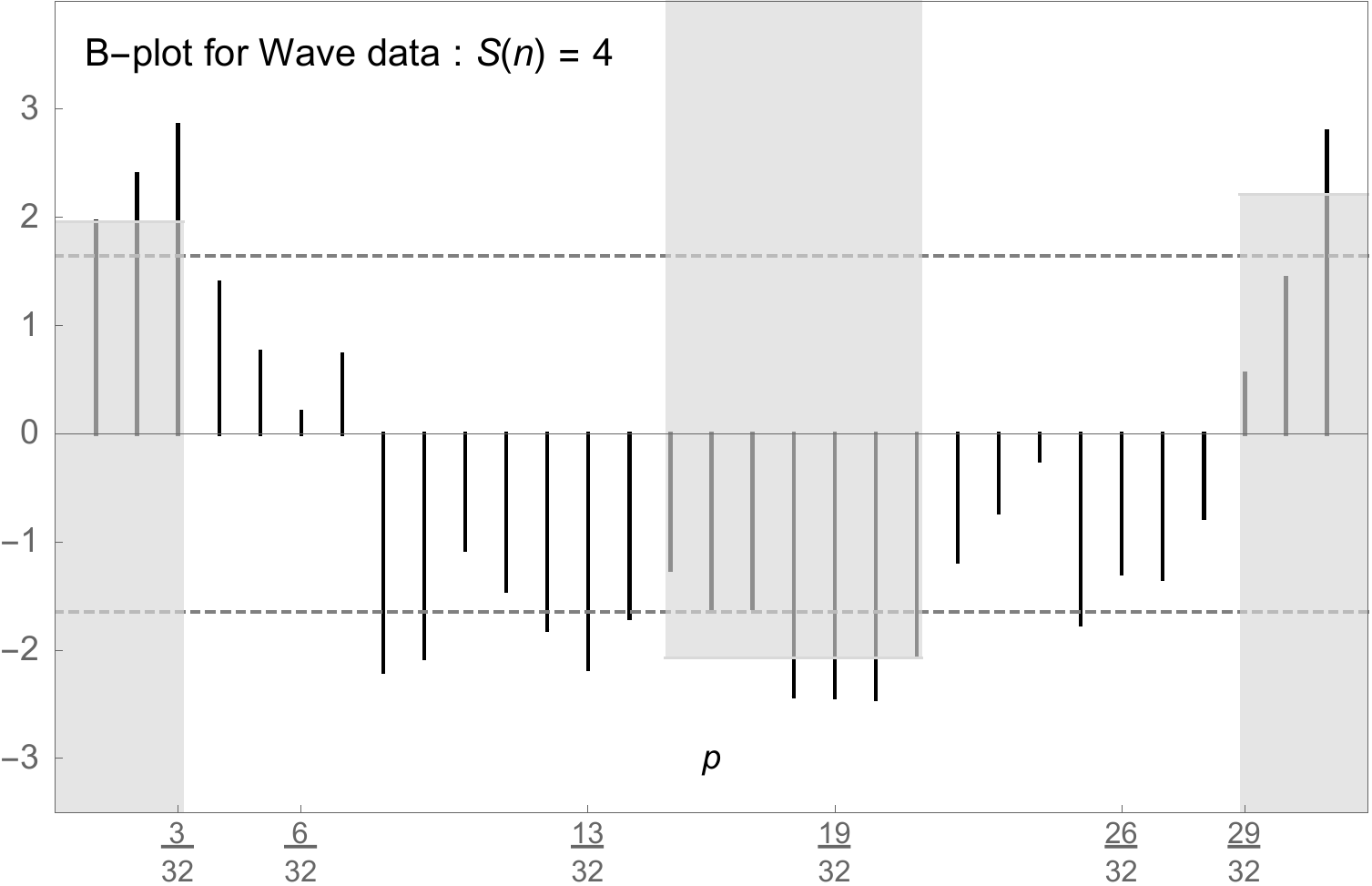}
\par\end{centering}
\centering{}\caption{\label{fig:Empirical-FC for the B=000026 D wave data set} B-plot
of $n^{1/2}\,\widehat{\textnormal{CC}}(p_{4,j};\tilde{\beta})$,
with $j\in\{1,,\ldots,31\}$, for the wave records data  ($n=66)$ in Bickel \& Doksum (1977). The dashed grey lines are located at $\pm1.645$
to identify the individual bars where some dissonance with $\mathbb{H}$: {\it Gaussianity}
occurs at one-sided level 5\%. The shaded grey stripes are the simultaneous $95 \%$ one-sided acceptance regions for subsets of bars \{1, 2, 3\}, \{15,\ldots, 21\} and  \{29, 30, 31\}.}
\end{figure}

%The value of $n^{1/2}\,\widehat{\textnormal{CC}}(p_{4,31};\tilde{\beta})$
%above the dashed line is consonant with the finding of Aldor-Noiman
%et al. (2013) of a fatter right tail than a Gaussian distribution.
%Assuming \textcolor{red}{from their previous work on this data set} that the meaning of their  term ``right tail''  relates to quantiles such that $p \geq 29/32 \approx 0.9$,  we can substantiate this
%by computing from (\ref{eq:p-value case 3}) $u(66,0.05; \{ 29,30,31 \}) = 2.21$. From Figure~\ref{fig:Empirical-FC for the B=000026 D wave data set}, because the bar at $p=31/32$ is above 2.21, this further supports the claim of Aldor-Noiman et al. (2013). 
%If we define similarly the left tail as below the  first decile, i.e. $p \leq 3/32$, we get $u(66,0.05; \{ 1,2,3 \}) = 1.96$. Again from Figure~\ref{fig:Empirical-FC for the B=000026 D wave data set}, all three bars in $\{ 1,2,3 \}$ are above this value,  in agreement with  the finding in Rosenkranz (2000)  regarding a thinner than Gaussian left tail.

To better appreciate the structure of the data,  Figure  \ref{fig:A.1 wave data for sn=6} presents the B-plot associated with the denser partition  $S(n) = 6$.  This figure shows that the overall shape of the B-plot is retained, while evidence of departures in the tails are better manifested. The question arises whether the conclusions regarding sets of bars hold for this new B-plot. We have recomputed $u(66,0.05,\{1,\ldots,12\})=2.27$,  $u(66,0.05,\{116,\ldots,127\})=2.57$ and  $\ell(66,0.05,\{58,\ldots,84\})=-2.43$ and represented the related simultaneous acceptance regions as shaded grey stripes. 
This shows that increasing $S(n)$ slightly changes the bounds, as should be expected, but that the previous observations remain unchanged.

By construction, the simultaneous acceptance regions are accurate up to the number of Monte Carlo replications, taken here as 100~000. However, one may inquire about the preciseness of the asymptotic  $\pm1.645$ individual bounds. In Figure  \ref{fig:A.1 wave data for sn=6} we added (the dotted red lines) the bounds obtained,  again from 100~000 Monte Carlo replications.  For finite samples,  the $n^{1/2}\,\widehat{\textnormal{CC}}(p_{s,j} \,;\,\tilde{\beta})$  have a discrete distribution. But interestingly,  even with the small sample considered here ($n=66$), the asymptotic 5\% bounds are sufficiently close to the exact values to be useful throughout the range $p\in (0,1)$, except perhaps near the outmost boundaries.

\begin{figure}[H]
\begin{centering}
\includegraphics[width = 0.75\linewidth]{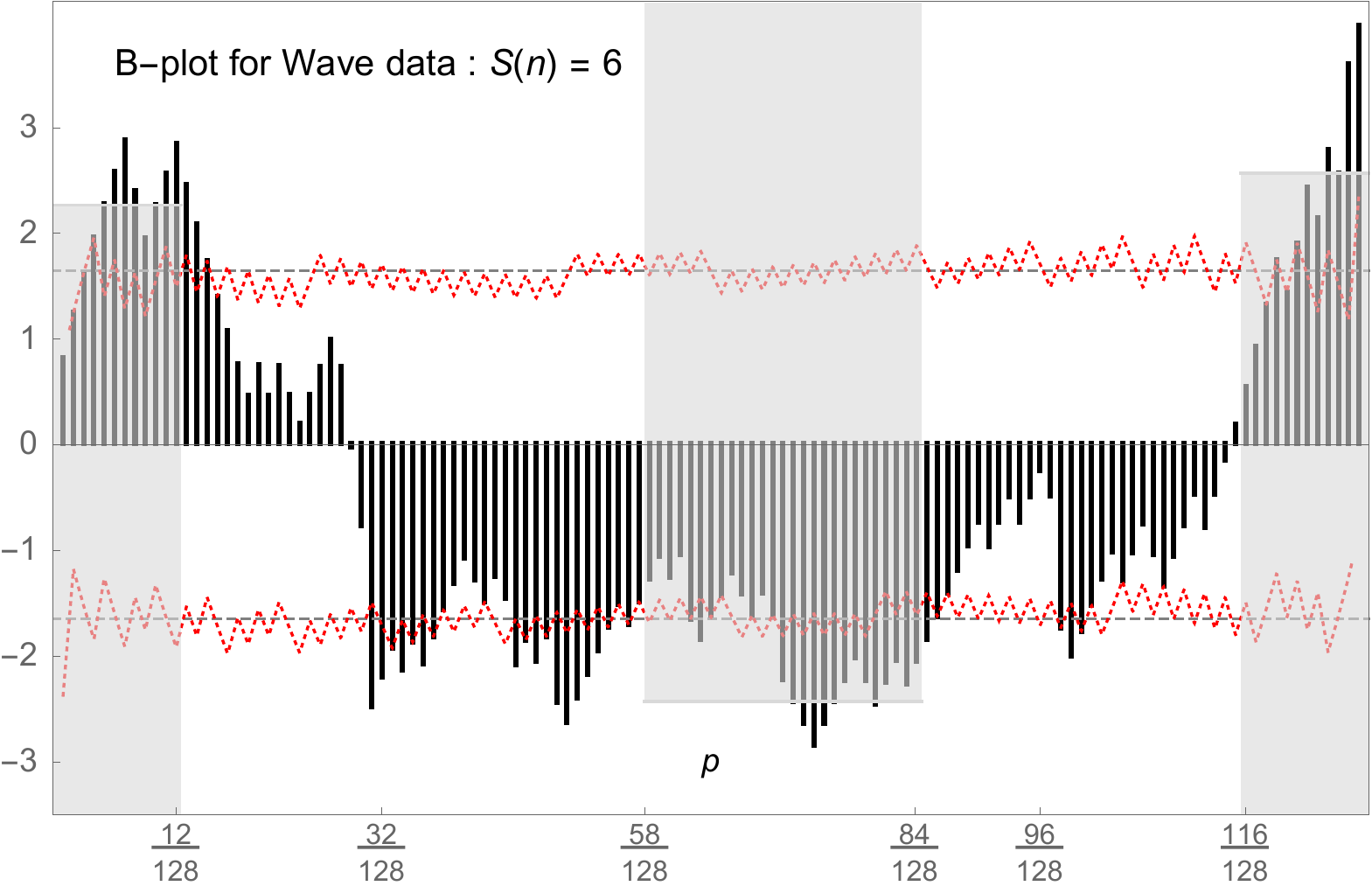}
\par\end{centering}
\centering{}\caption{\label{fig:A.1 wave data for sn=6} B-plot
of $n^{1/2}\,\widehat{\textnormal{CC}}(p_{6,j};\tilde{\beta})$,
with $j\in\{1,,\ldots,127\}$, for the wave records data ($n=66)$ in Bickel \& Doksum (1977). The dashed grey lines are located at $\pm1.645$
to identify the individual bars where some dissonance with $\mathbb{H}$: {\it Gaussianity}
occurs at one-sided level 5\%. The shaded grey stripes are the simultaneous $95 \%$ one-sided acceptance regions for subsets of bars \{1,\ldots, 12\}, \{58,\ldots, 84\} and  \{116,\ldots, 127\}. The dotted red lines are the one-sided (upper and lower)  $95\%$ acceptance intervals for an individual bar, as obtained by the Monte Carlo method.}
\end{figure}

A B-plot  can also provide useful  information regarding which test statistic could be more profitably applied to assess overall compatibility between the data and the model.
For example, the above analysis exhibits substantial disagreements  between the data and the model in the tails.  Now, much evidence (see Milbrodt \& Strasser, 1990; \'Cmiel, Inglot \& Ledwina, 2020;  Inglot,  2020 and references therein) have been unearthed showing that in such circumstances, the classical Kolmogorov-Smirnov (KS) test is weak: for the wave data we get a $p-$value of 0.06. B-plots present weighted distances between an estimated empirical process and the model CDF. This weighting rescales the differences appearing in the KS statistic, thus creating a comparable  scale, under $\mathbb{H}$, over the whole range of $p$. In particular, test statistic $\mathcal{M}_{127}(\tilde\beta)$, which can be considered as a weighted variant of the KS statistic, leads to a $p-$value of  0.004, comparable with AD, SW and BCMR,  thus removing the weakness of the classical KS solution. In contrast, in situations where the B-plot  shows that most of the discrepancies occur in the central range of quantiles, such classical tests can be adequate tools,  see the next section.

The above ranges for $p_{S(n),j}$ can be translated into the original scale of the data via ~$\tilde{\beta}_{2}\Phi^{-1}(p_{S(n),j})+\tilde{\beta}_{1}$; see Appendix~A.1.

\subsubsection*{3.5.2~The tephra data }

 % Tephra data

We consider the tephra data ($n=59$) analyzed in Bowman \& Azzalini (1997, Section~2.5).
We apply the transformation to the logistic scale (i.e. $X=\textrm{log} \big(Y/(100-Y)\big)$ as done by these authors. For this data, we  find $\tilde\beta_1=\bar{X}=-1.77$,  $\tilde\beta_2= S = 0.056$. Here we reverse the order in which our tools were applied in the previous example and first look at the B-plot for this data set, which appears in  Figure \ref{fig:Empirical-FC- tephra data} for $S(n) = 4$ and $S(n) = 6$.  A few bars are unexpectedly large in case $\mathbb{H}$ is true, in the central region  $p \in(13/32,19/32)$.  This is substantiated by computing (for $S(n) = 4$) $u(59,0.05; \{ 13,\ldots,19 \}) = 2.10$. 

\begin{figure}[H]
\begin{centering}
\includegraphics[width =\linewidth]{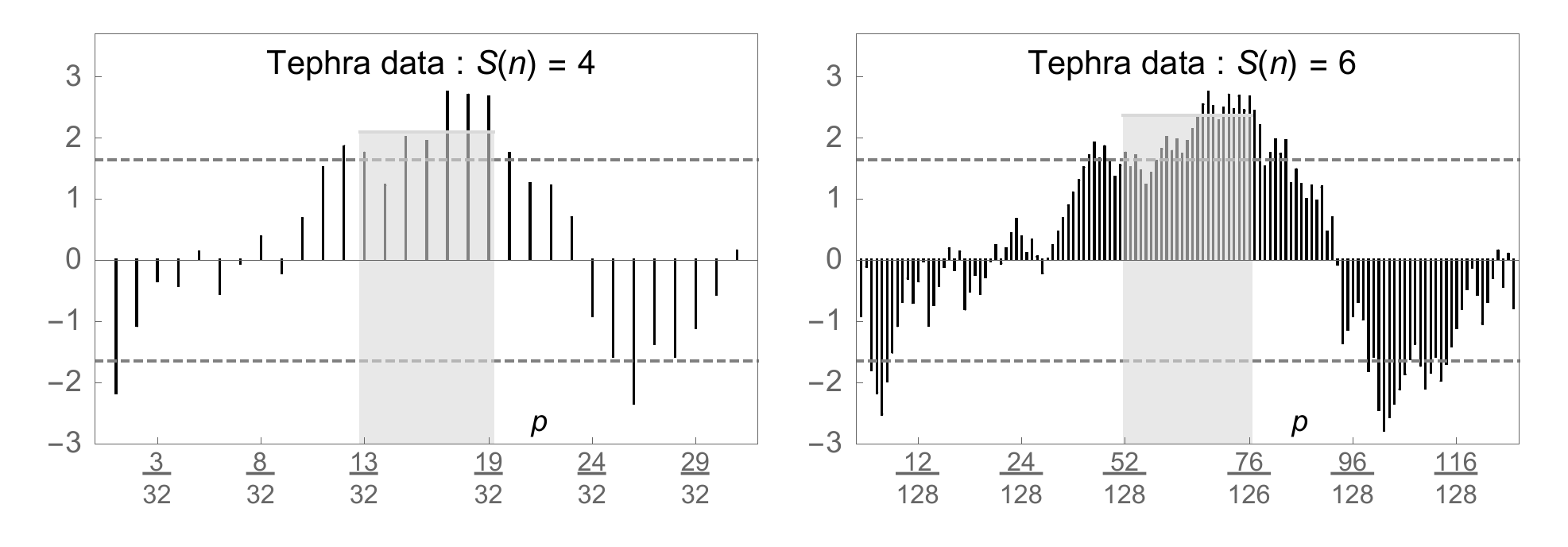}
\par\end{centering}
\caption{\label{fig:Empirical-FC- tephra data} B-plot of the values
of $n^{1/2}\,\widehat{\textnormal{CC}}(p_{S(n),j};\tilde{\beta})$, with $S(n)=4$ and  $S(n)=6$
for the tephra data ($n=59$)
from Bowman \& Azzalini (1997). The dashed grey lines are located
at  $\pm1.645$ to identify the bars where dissonance with $\mathbb H$: 
{\it Gaussianity} occurs. The shaded grey stripe is  the simultaneous $95 \%$ one-sided acceptance region for bars in the indicated deciles.}
 \end{figure}

As stated in Section~3.5.1, there is a vast amount of literature providing some indications as to which test is more efficient in some given situations. In particular, when relatively large departures occur near the centre of the data, the classical solutions, e.g. KS, CvM and AD, have been shown to be efficient. In particular, the AD test works well as it can also detect allocations of moderate portions of mass towards the tails. For related discussions, see Janssen (2000), Inglot, Kallenberg \& Ledwina (2000), \'Cmiel, Inglot \& Ledwina (2020) along with its Supplementary Information and references therein. To substantiate this evidence here, we have applied the tests SW and BCMR  which do not  reject,  with $p-$values for 0.13 of SW and 0.12 for BCMR.   The test based on $\mathcal{P}_{\tilde{Q}(\alpha)}(\tilde{\beta})$, with $S(n) = 4$ and using the same constants as previously, 
yields an observed value of  $\mathcal{T}_{n}$ of 1.79, leading to $\tilde{Q}(\alpha) $ = 1 with  $\mathcal{P}_{\tilde{Q}(\alpha)}(\tilde{\beta})=3.78$,  to be compared to 
a $5\%$ critical level of 10.47.  Thus the null hypothesis of Gaussianity is also not rejected at the $5\%$ level. However, with the KS test, we get a $p-$value of 0.051 while the AD test yields a $p-$value of 0.03. This shows that the observation of the B-plot can provide some clues as to what test, here one of the classical solutions, should be subsequently applied to formally detect global departures between the data and the model.

 \subsubsection*{3.5.3~The PCB data }
We consider the PCB data set of Risenbrough  ($n=65$) recalled in Thas (2010, p. 5) and pertaining to the concentration of the chemical PCB (polychlorinated biphenyl) in the yolk lipids of Anacapa (pelican) birds. The data has been thoroughly studied by the author using several graphical methods and one of his conclusions, based on an estimated comparison density,  is (Thas, 2010, p. 73)  ``the plot suggests weakly that the frequency of PCB concentrations is smaller than expected under the hypothesis of normality\,''. 

For this data, we  have $\tilde\beta_1=\bar{X}=210.0$ and $\tilde\beta_2= S =72.26$. Here again, we first look at the B-plot for this data set, which appear in  Figure \ref{fig:PCB data sn=4} and is plotted using both $S(n) = 4$ (with $\ell(65,0.05; \{ 16,\ldots, 25 \}) = -2.27$) and $S(n) = 6$ (with $\ell(65,0.05; \{ 64,\ldots,100 \}) = -2.49$).

\begin{figure}[H]
\begin{centering}
\includegraphics[width =\linewidth]{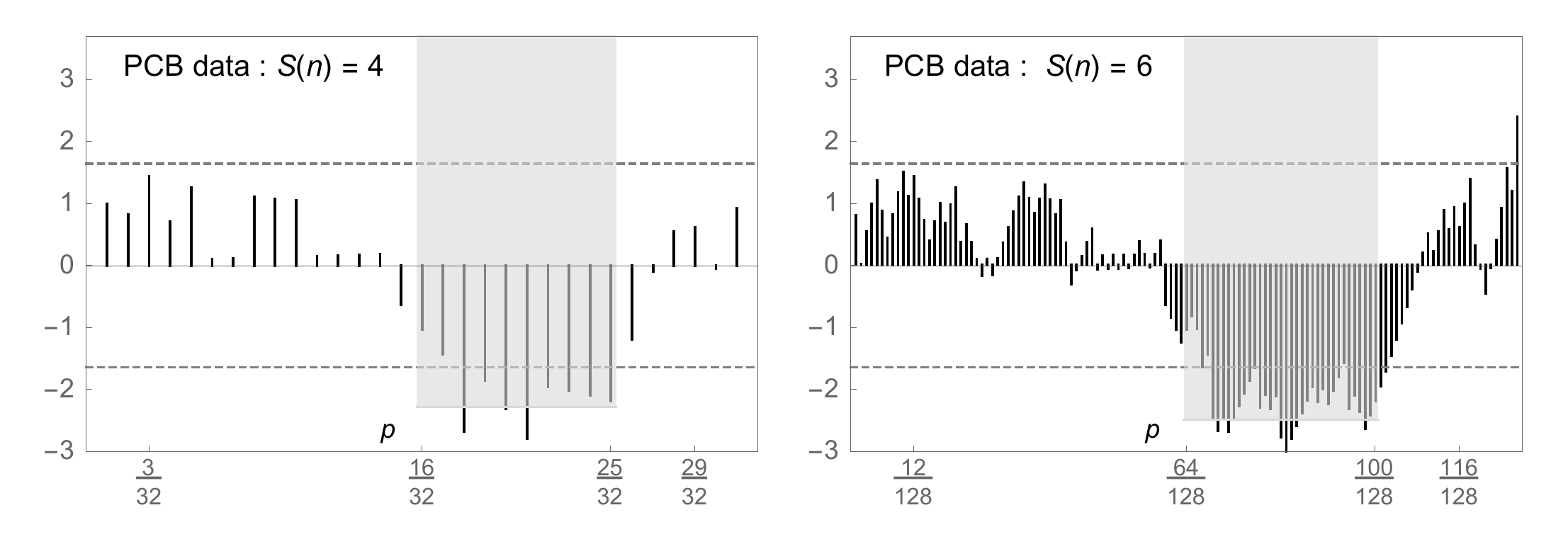}
\par\end{centering}
\caption{\label{fig:PCB data sn=4} B-plot of the values
of $n^{1/2}\,\widehat{\textnormal{CC}}(p_{S(n),j};\tilde{\beta})$, with $S(n)=4$ and  $S(n)=6$
for the PCB data ($n=65$). The dashed grey lines are located
at  $\pm1.645$ to identify the bars where dissonance with 
$\mathbb H$: 
{\it Gaussianity} occurs. The shaded grey stripes are  the simultaneous $95 \%$ one-sided acceptance region for bars \{16,\ldots, 25\} and  \{64,\ldots, 100\}. }
 \end{figure}
 
 The left half of the distribution appears concordant with the Gaussian hypothesis while the data seems more concentrated toward the centre on the right side.  In contrast to previous examples, there seems to be little guidance formulated in the literature suggesting that one of the classical  tests for this problem may be  preferable in this case. In particular, we get for KS a $p-$value of 0.057, 0.056 for SW,  and 0.057 for AD, all near but above the 0.05 threshold. The BCMR test barely rejects Gaussianity with a $p-$value of 0.049. This is a situation where an adaptive approach such as Thas's (2010, p. 120) data driven smooth test based on Hermite polynomials could be useful. Such a test yields a $p$--value of 0.0325,  which leads to  rejection of the Gaussian hypothesis. However, the author is unable the derive from the test's components any insight about what may have caused rejection. Our data driven test based on $\mathcal{P}_{\tilde{Q}(\alpha)}(\tilde{\beta})$, with $S(n) = 4$ and using the same constants as previously,  yields an observed value of  $\mathcal{T}_{n}$ of 2.64, leading to $\tilde{Q}(\alpha) $ = 31 with  $\mathcal{P}_{\tilde{Q}(\alpha)}(\tilde{\beta})=56.74$,  to be compared to  a $5\%$ critical level of 10.47.  Thus the null hypothesis of Gaussianity is here rejected at the $5\%$ level, a conclusion emhanced by the above knowledge derived from the B-plot about the discrepancies with the model.

 \subsubsection*{3.5.4~The smiling baby data revisited}
 
Here, we revisit the smiling baby data set of Section~2.1, normalized to $[0, 1]$. The B-plot for this data with $S(n)=4$
appears in Panel 2) of Figure 1. The B-plot associated with $S(n)=6$ appears in Figure \ref{fig:smiling data sn=6}. To obtain more precise insights regarding some sets of adjacent bars in the B-plot, the variants of $u(n,\alpha, \{ r,\ldots,s \})$ and $\ell(n,\alpha, \{ r,\ldots,s \})$ adapted to a simple null hypothesis, with $\widehat{\textnormal{CC}}(p_{S(n),j})$ in place of $\widehat{\textnormal{CC}}(p_{S(n),j}),\tilde{\beta})$ as explained in Section~3.3, could be computed. However, this is unnecessary here because all individual bars are well within $\pm 1.645$ and no reason emerges to reject uniformity anywhere.

To validate this visual assessment,  we  apply the test based on $\mathcal{P}_{R(\alpha)}$ with $S(n)=6$ and $\alpha=0.05$. 
The observed value of $\mathcal{M}_{n}$ is 1.59, leading to $R(\alpha)$ = 1 with  $\mathcal{P}_{R(\alpha)}$ = 0.16, to be compared to 
a $5\%$ critical level of 133.9. Thus we clearly do not reject the null hypothesis of uniformity at the $5\%$ level. This is in agreement with the AD ($p-$value = 0.63) 
and BJ   ($p-$value = 0.85) tests.

\begin{figure}[H]
\begin{centering}
\includegraphics[width =0.70\linewidth]{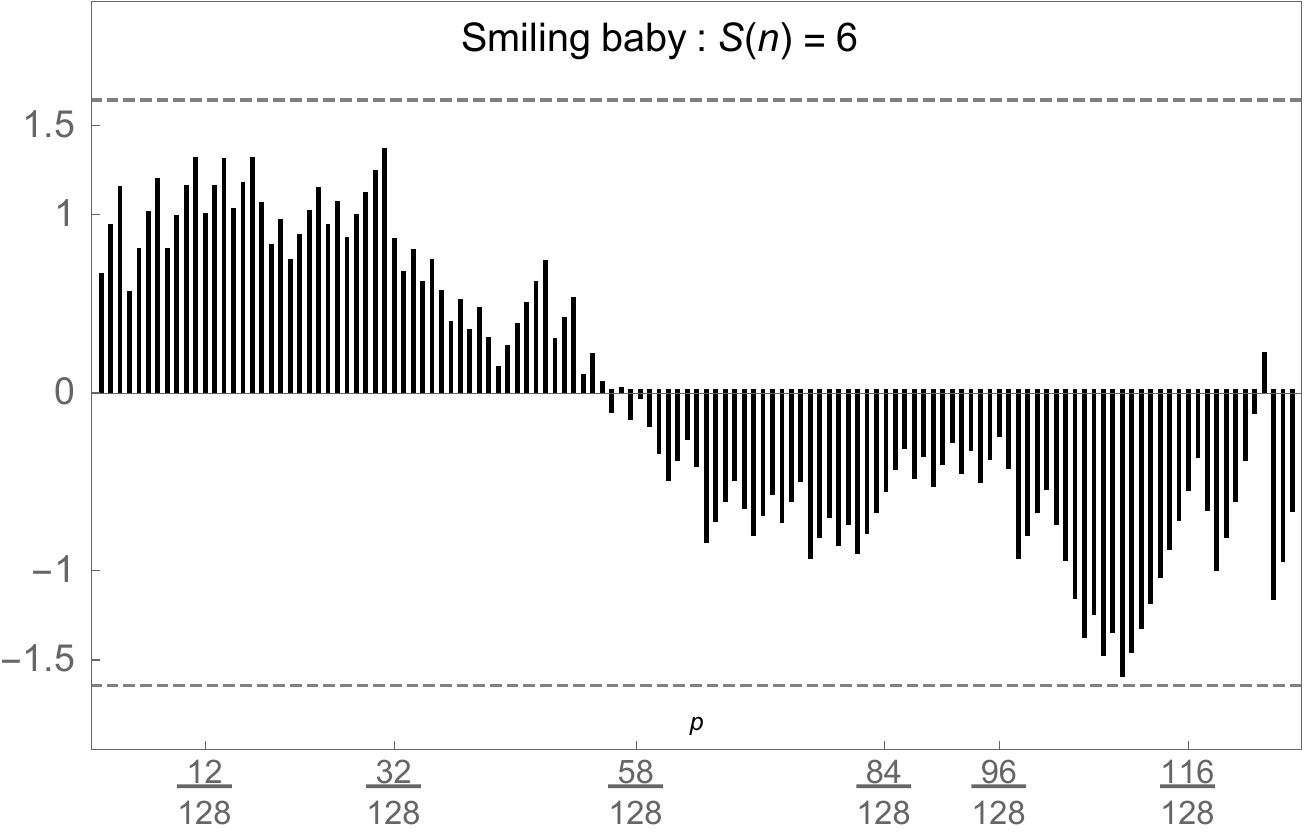}
\par\end{centering}
\caption{\label{fig:smiling data sn=6} B-plot of the values
of $n^{1/2}\,\widehat{\textnormal{CC}}(p_{S(n),j})$, with $S(n)=6$
for the smiling baby data ($n=55$). The dashed grey lines are located
at  $\pm1.645$ to identify the bars where dissonance with uniformity occurs.} 
 \end{figure}

For these data, we  have $\tilde\beta_1=\bar{X}=0.50$, the median is 0.47 and $\tilde\beta_2= S = 0.26$. From these values, one may wonder whether tests tailored to detect a shift to the left of the median or a smaller dispersion would  improve on the above general-purpose GoF procedures, i.e. AD and BJ. The one-sided signed rank test yields a  $p-$value of 0.45 while the one-sided Brown-Forsythe test has a $p-$value of 0.15.
These extra results further support the claim of previous authors (see Bhattacharjee \& Mukhopadhyay, 2013) that this data could well be uniformly distributed.

%% ############    Discussion  ################

\section{Discussion\label{sec:Section : Discussion}}

The present work proposes an approach for model validation based on
the pair (B-plot, $\mathcal{P}_{R(\alpha)})$
or (B-plot, $\ensuremath{\mathcal{P}_{\tilde{Q}(\alpha)}}(\ensuremath{\widetilde{\beta}})$)
that appears to be a good compromise to the two routes presented in
the Introduction. In the case of a Gaussian null model, with known as well as unknown parameters, our test statistics
are powerwise competitive with some of the best solutions proposed
in the literature, while the B-plot and related acceptance regions offer an enhanced 
assessment, with respect to PP or QQ plots, as to where the data deviate
from the contemplated model.  The technical details here are confined to
the Gaussian null model. But existing evidence and earlier experience allows 
to expect that our approach can be extended to other nonparametric problems, general location-scale families and
several more complex models. 
In particular, other $\sqrt{n}$-consistent
estimators than the MLE  could be covered by adjusting the  \textnormal{CC} curve, the test statistic and the data driven selection procedure. 
 For illustration of different situations where data driven smooth tests, using  other systems of functions than the present $\{h_{s,j}(\cdot)\}$, work well see Kallenberg \& Ledwina (1999),  Pe\~{n}a (2003), Ducharme \& Fontez (2004), Ducharme \& Lafaye de Micheaux (2004, 2020), Inglot \& Ledwina (2006), Bissantz, Claeskens, Holzmann \& Munk (2009), Escanciano \& Lobato (2009), Janic \& Ledwina (2009), Wang \& Qu (2009),  Wy{\l}upek (2010, 2021) and Thas, Rayner \& de Neve (2015). However, it should be noted that in such situations, the constructions are more involved as a rule and additional technical work is needed. In return, both empirical and theoretical studies show that well-calibrated data driven tests are only slightly less powerful than classical solutions applied in their most favourable situations, while otherwise having unrivalled sensitivity to a large spectrum of important alternatives. 

We emphasize that the main motivation for this work is to propose an approach that can help in understanding the structure of the data at hand, to allow investigating why a null hypothesis has been rejected, to detect and describe some local discrepancies, and to provide some evidence in which sense and how reliable such an approach can be. 
In particular, we consider such endeavors as useful additions that better enshrine the modeling process in a more constructive iterative loop, as  pointed out in the first paragraph of the present Introduction. We confine our work here to the case of classical goodness of fit testing  but note that questions of this kind are increasingly discussed in recent literature on several different testing problems. For an illustration see Kim et al. (2019), Zhang (2019), Algeri (2021), Xiang et al. (2023)  and the references therein.

In Sections~2.5 and 3, we have qualified our alternatives, comprising a range of shape of \textnormal{CC}'s, as ``carefully'' selected.
In many simulation studies
about the empirical power of GoF tests, the alternatives are taken among
broad categories such as symmetric, asymmetric, etc., often building up on previous simulations by adding
some new ``interesting'' alternatives. In addition, in summarizing
their results, many authors base their final recommendations on some
averaging of the obtained powers over the alternatives investigated.
However, such categories are mostly related to shape of densities,
which may not be well adapted to departures related to other characteristics
that some GoF tests can detect with greater power than density differences.
Thus such averaging can introduce bias in these recommendations, which are often of the form ``this test is good at detecting such type of departures'' with,
in many cases, departures pertaining to asymmetry and large or short tails.  However, at the beginning of the modelling process, a user has often limited
knowledge about the plausible alternatives to the null model. Hence such recommendations are of little help in choosing a GoF test appropriate
to his problem and this will often lead to the use of a test  based solely on its popularity. This is not good science. Here, examination of the B-plot, as in the examples of  Section~ 3.5, allows  acquiring such knowledge and 
decide whether one can use with some confidence a classical solution or would be better off going through the trouble of considering a much more computationally
expensive data driven test, as the ones of the present work. 

\textcolor{black}{\section*{Acknowledgements} The authors would like to thank the AE for his/her handling of the manuscript and some useful suggestions. We are also very grateful to  an anonymous referee for constructive contributions that greatly improved the readability of the paper. Finally the authors would like to thank Professor Pierre Lafaye de Micheaux for his help with the installation of the codes performing the computations of the paper on the repository site GITHUB.}

% %  ############    bibliographie  ################

%\bibliographystyle{imsart-nameyear}\bibliographystyle{biometrika.bst}

\renewcommand\refname{References for Sections 1 to 4}

\newpage
%   ############    Appendix A :Some examples   ################

\appendix

\vspace*{4pt}

%\appendixtwo
\section{More real data examples}
%\vspace*{5pt}

%\setcounter{figure}{0}
%\renewcommand{thefigure}{C\arabic{figure}}

%\makeatletter
%\renewcommand{\thefigure}{C\arabic{figure}}
%\makeatother

%\subsection*{Real data examples}

This appendix  contains details regarding three examples that show the  information that can derive from the tools of the paper.

\subsection*{A.1.~~The wave record data revisited} 

The B-plots in Section 3.5 are expressed as functions of the quantiles $p$. A variant B-plot can be produced that relates more directly to the original data. To this end, 
set $\tilde q_{s,j} =\tilde q_{s,j}(p_{s,j}) =  \tilde{ \beta_2} \,\Phi^{-1}(p_{s,j}) + \tilde \beta_{1}$, which represents the estimated $p_{s,j}$ quantile of the null distribution. With this notation, set
\begin{align*}
\widetilde{\textnormal{CC}}(\tilde{q}_{s,j} \,;\,\tilde{\beta}) & = \frac{\Phi\big((\tilde{q}_{s,j} -\tilde\beta_{1})/\tilde\beta_2\big)- \hat{F}_{n}(\tilde{q}_{s,j})}{\sigma_{s,j}}\\
& =  \frac{p_{s,j} - \hat{F}_{n}(\tilde{ \beta_2} \,\Phi^{-1}(p_{s,j}) + \tilde \beta_{1})}{\sigma_{s,j}},
\end{align*}
where $\hat{F}_n(\cdot)$ is the ordinary empirical CDF of the sample. Given $s$, this variant of the B-plot, noted $\textrm{B}_q$-plot, is obtained by plotting the $\widetilde{\textnormal{CC}}(\tilde{q}_{s,j} \,;\,\tilde{\beta})$ against the $\tilde q_{s,j}~ (j=1,\ldots, d(s))$. Figure \ref{fig:A.1.2 wave data for sn=5}  shows such a graph for the wave data of Section 3.5.1 with $S(n)=5$ and the related acceptance regions. In  particular, one can see that the seven data points greater than  $ \tilde {q}_{5,58}= 6.9$ are more dispersed to the right than expected under the null hypothesis.

\renewcommand{\thefigure}{A.1}

\begin{figure}[H]
\begin{centering}
\includegraphics[width = 0.75\linewidth]{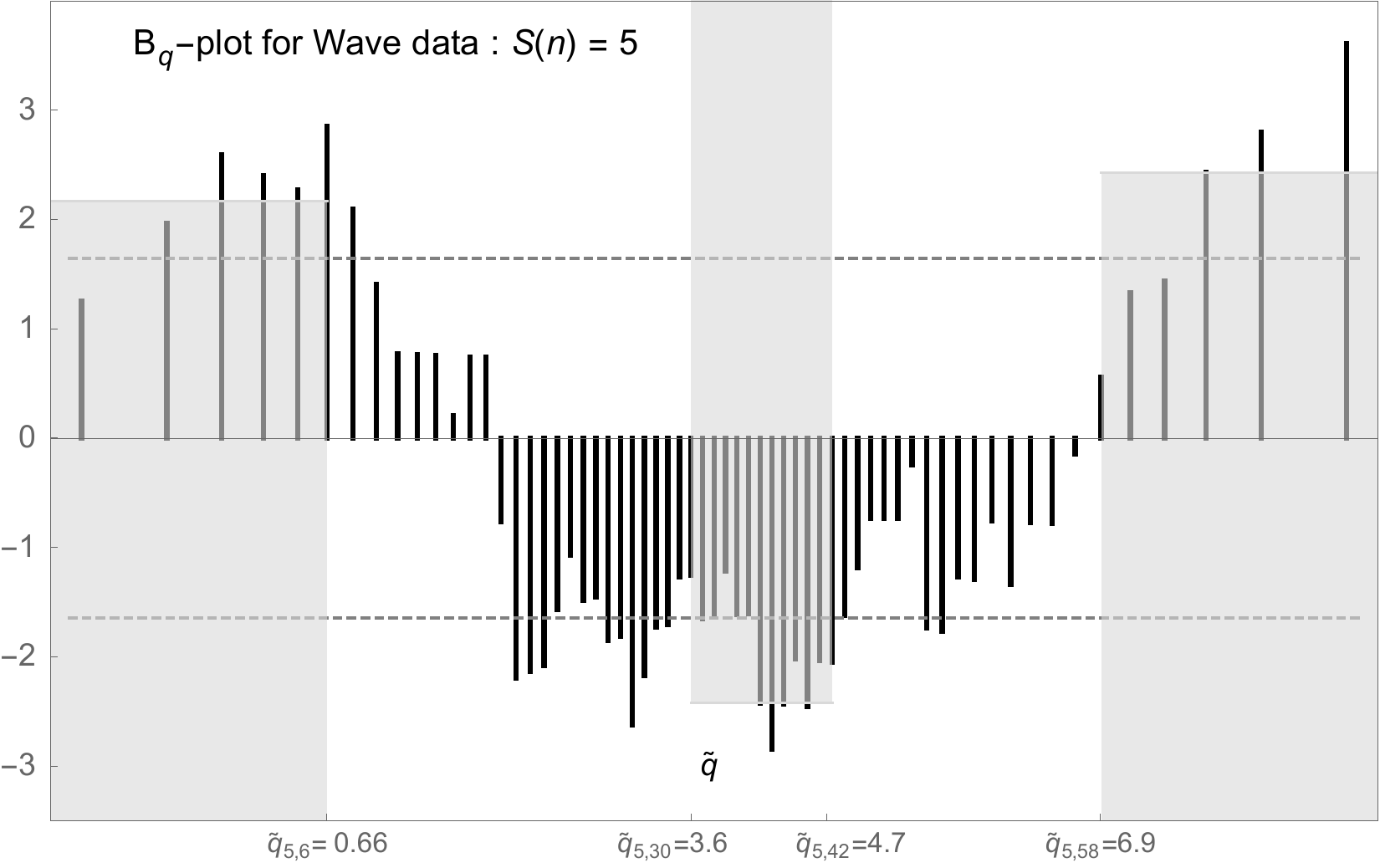}
\par\end{centering}
\centering{}\caption{\label{fig:A.1.2 wave data for sn=5} $\textrm{B}_q$-plot
of the values of $n^{1/2}\,\widetilde{\textnormal{CC}}(\tilde{q}_{5,j};\tilde{\beta})$ plotted against the estimated quantiles $\tilde q_{5,j}$
with $j\in\{1,\ldots,63\}$, for the wave records data ($n=66)$ in Bickel \& Doksum (1977). The dashed grey lines are located at $\pm1.645$
to identify an individual bar where some dissonance with $\mathbb{H}$: {\it Gaussianity}
occurs at one-sided level 5\%. The shaded grey stripes are the simultaneous $95 \%$ one-sided acceptance regions for subsets of bars \{1,\ldots, 6\}, \{30,\ldots, 42\} and  \{58,\ldots, 63\}.}
\end{figure}

\newpage

% Open/Closed book data

\subsection*{A.2.~~The open/closed book examination data}
Consider  the open book /closed book examination data set in Mardia, Kent \& Bibby (1979) pp. 3--4, 
which gives the marks
of a group of \emph{n} = 88 students in Mechanics, Vectors, Algebra,
Analysis, and Statistics. The marks in Statistics, Vectors and Analysis
were analyzed in Ducharme \& Lafaye de Micheaux (2020) who
rejected a trivariate multinormal
distribution. Here we revisit the marks for Analysis. 

For this data set, we  have $\tilde\beta_1=\bar{X}=46.68$ and $\tilde\beta_2= S = 14.76$.
We have applied the test based on $\mathcal{P}_{\tilde{Q}(\alpha)}(\tilde{\beta})$
with $S(n)=4$ and $\alpha=0.05$. 
The observed value of $\mathcal{T}_{n}$ is 5.15, leading to $\tilde{Q}(\alpha) $ = 31 with  $\mathcal{P}_{\tilde{Q}(\alpha)}(\tilde{\beta})$ = 155.12, to be compared to 
a $5\%$ critical level of 10.44, interpolated from Table \ref{Tab5} in  Appendix~B. Thus we reject the null hypothesis of Gaussianity at the $5\%$ level.
The $p-$values for AD, SW, BCMR are 0.0001 for AD,  0.0001 for SW  and  0.001 for  BCMR. Thus these tests also reject the Gaussian model at level $5\%$.

\renewcommand{\thefigure}{A.2}

\begin{figure}[H]
\begin{centering}
\includegraphics[width = \linewidth]{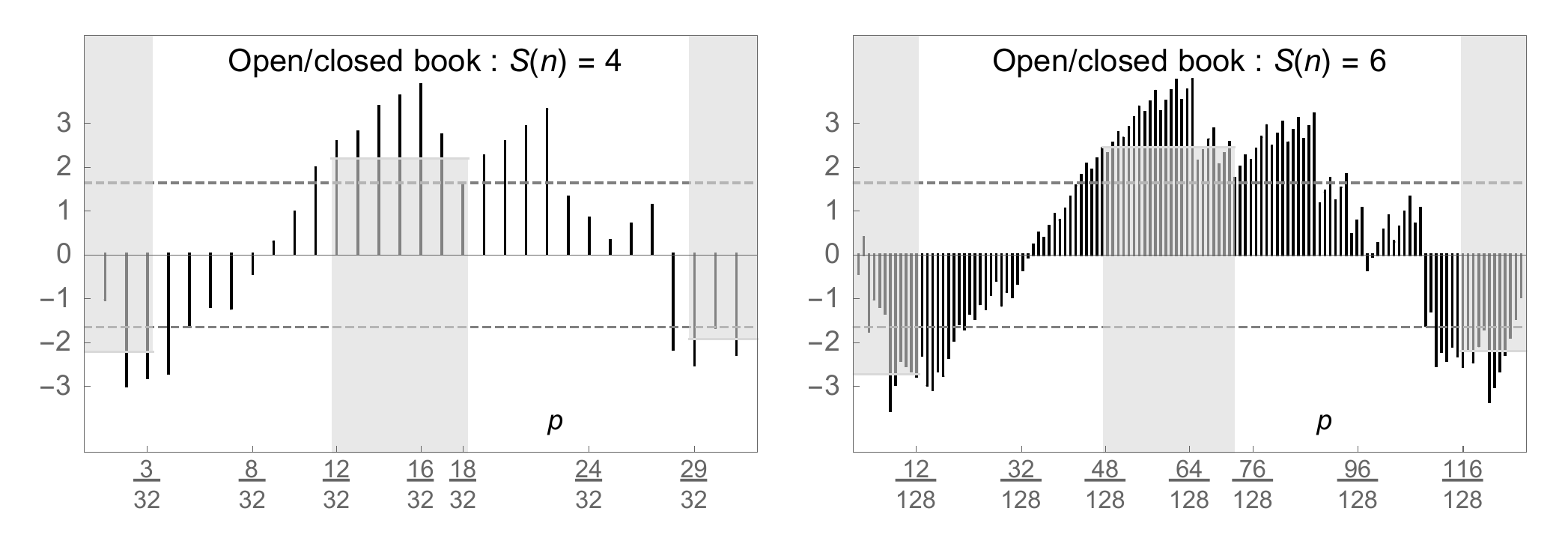}
\par\end{centering}
\caption{\label{fig:bar display Open/closed book}B-plot 
of $n^{1/2}\,\widehat{\textnormal{CC}}(p_{S(n),j};\tilde{\beta})$ with $S(n)=4$ and  $S(n)=6$
for the marks in analysis in the open/closed book data ($n=88$)
from Mardia, Kent \& Bibby (1979). The dashed grey lines are located
at $\pm1.645$ to identify a bar where dissonance with 
$\mathbb H$: 
{\it Gaussianity} occurs. The shaded grey stripes are the simultaneous $95 \%$ one-sided acceptance regions for subsets of bars in the first and last deciles and the slightly decentred  central
part $p \in (0.38,0.56)$.}
\end{figure}

On the B-plot of Figure \ref{fig:bar display Open/closed book}, the left tail of the distribution seems heavier than the Gaussian, while the right one could be thinner,
thus indicating some asymmetry to the right of the distribution with less mass at the centre. We find for $S(n)=4$,   $\ell(88,0.05; \{ 1,2,3 \}) = -2.21$, $\ell(88,0.05; \{ 29,30,31 \})$ = $-1.92$ and, for similar reasons as in Section 3.5,  the slightly decentred  (to the left)   $u(88,0.05; \{ 12,\ldots,18 \}) = 2.21$. After drawing the related acceptance regions, we can observe that, in all regions, at least one bar goes beyond these acceptance regions, thus supporting the above claims. 
\renewcommand{\thefigure}{A.2}

\newpage

\subsection*{A.3.~~The cystine data} 
Consider the cystine content of grade 5 yellow corn. The data  ($n=106$) appear in Gan, Koehler \& Thompson (1991). In
their work  they use graphical methods to infer, from the shape of
the PP plot, that the Gaussian distribution does not provide an adequate
model. However after several attempts, they conclude that no other
model they have investigated leads to a clearly better alternative
to the Gaussian. 
\vspace*{2pt}

For this data set, we  have $\tilde\beta_1=\bar{X}=0.09$ and $\tilde\beta_2= S = 0.014$.
We apply the test based on $\mathcal{P}_{\tilde{Q}(\alpha)}(\tilde{\beta})$
with $S(n)=4$ and $\alpha=0.05$.
The observed value of  $\mathcal{T}_{n}$ is 4.20, leading to $\tilde{Q}(\alpha) $ = 31 with  $\mathcal{P}_{\tilde{Q}(\alpha)}(\tilde{\beta})$ = 118.52 to be compared to 
a $5\%$ critical level of 10.42. Thus the null hypothesis of Gaussianity is rejected at the $5\%$ level.
The  tests  AD, SW and BCMR similarly reject at the $5\%$ level  with the $p-$values for AD : 0.0001, SW : 0.003 and BCMR : 0.005.

\renewcommand{\thefigure}{A.3}

\begin{figure}[H]
\begin{centering}
\includegraphics[width =\linewidth]{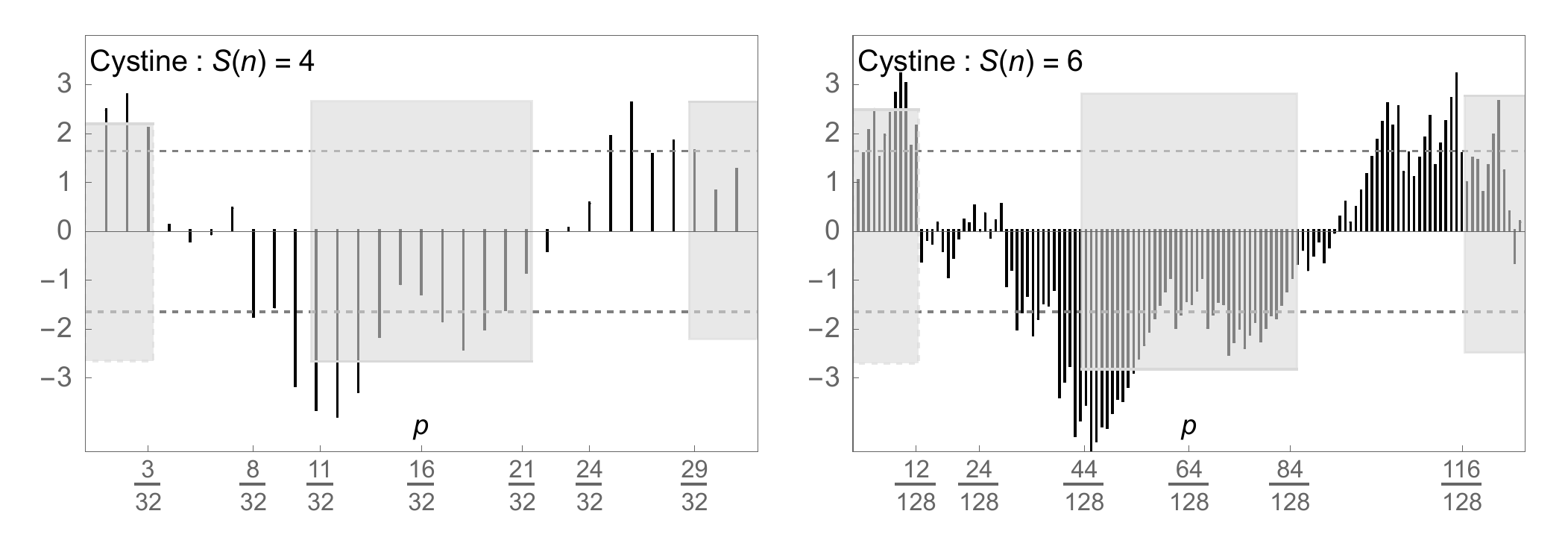}
\par\end{centering}
\caption{\label{fig:Empirical-FC- Cystine data}B-plot of the values
of $n^{1/2}\,\widehat{\textnormal{CC}}(p_{S(n),j};\tilde{\beta})$ with $S(n)=4$ and  $S(n)=6$
for the cystine data ($n=106$) from  Gan, Koehler \& Thompson (1991). The dashed grey lines are located
at $\pm1.645$ to identify a bar where dissonance with $\mathbb H$: 
{\it Gaussianity} occurs. The shaded grey rectangles are the two-sided $95 \%$ acceptance regions for the subsets of bars in the first and last decile and the middle tier of the distribution.}
\end{figure}

\textcolor{black}{In view of the limited previous information regarding this data set, and to obtain some insights about the type of departures from Gaussianity suggested by the data,  it may appear appropriate to consider two-sided acceptance regions for a few subsets of bars in the  B-plots. }Here we consider the subsets of bars in the first and last decile (the tails) and the middle tier of the distribution.  B-plots of the $n^{1/2}\,\widehat{\textnormal{CC}}(p_{S(n),j};\tilde{\beta})$ for $S(n) = 4$ and $S(n) = 6$, along with the  $\pm1.645$ critical values (see Figure~\ref{fig:Empirical-FC- Cystine data}) suggest a left tail thinner than the Gaussian but a right tail more consonant with the Gaussian.  For $S(n)=4$ (resp. $S(n)=6$) we get $\ell(106,0.025; \{ 1,2,3 \}) = -2.65$ (resp. $-2.70$ using $\mathcal{J}= \{ 1,\ldots,12 \}$)   and $u(106,0.025; \{ 1,2,3 \}) = 2.20$ (resp. $2.49$),  supporting the claim about the left tail. For the right tail, we find $\ell(106,0.025; \{ 29,30,31 \}) = -2.19$ (resp. $-2.48$  using $\mathcal{J}= \{ 116,\ldots,127 \}$) and $u(106,0.025; \{ 29,30,31 \}) = 2.65$ (resp. $2.77$) and no discrepancy seems to occur there. As for the central tier part, we  consider the quantiles in the  range $0.33  \approx 11/32 \leq p \leq  21/32  \approx 0.66$ and compute \textcolor{black}{  $\ell(106,0.025; \{ 11,\ldots,21 \}) = -2.66$ (resp. $-2.82$ using $\mathcal{J}= \{ 44,\ldots,84 \}$), $u(106,0.025; \{ 11,\ldots,21 \}) = 2.65$ (resp. $2.81$). }The data seems indeed more abundant in this region.

%   ############    Appendix B :Practical aspects   ################
\newpage

\section{Details and recommendations  on the implementation of our tools}
\setcounter{table}{0}
\renewcommand{\thetable}{B.\arabic{table}}

The practical implementation of our tests requires the choice of $S(n)$ and, depending on whether $\beta$ is known (the case of $\mathbb{H}_{0}$ of Section 2) or must be estimated (the case of $\mathbb{H}$ : $\Phi\big(({x-\beta_{1})/}{\beta_{2}}\big)$ of Section 3.4),  the computation of $a(n,\alpha)$ or $a(n,\alpha;\tilde\beta)$. In addition, our tests require  the values $m(n,\alpha)$ or $t(n,\alpha)$ for  the oracles  $\mathcal{M}_{D(n)}$ or $\mathcal{T}_{n}$ as well as the critical values $c(n,\alpha)$ for $\mathcal{P}_{{R}(\alpha)}$  or  $\tilde{c}(n,\alpha)$ for $\mathcal{P}_{\tilde{Q}(\alpha)}(\tilde\beta)$. Here we give some details about these values for the cases where  $4 \leq S(n) \leq 6$ and for values of $n$ in the range ${50,\ldots,500}$, which should cover many situations encountered in practice. Linear interpolations can be used in between entries of the tables.

The values of  $a(n,\alpha)$ and $a(n,\alpha;\tilde\beta)$  are rather stable as functions of $n$. In particular, one can take $a(n,0.1) = 2.59 $, $a(n,0.05) = 3.31 $ while $a(n,0.10;\tilde\beta)  = 2.53$ and $a(n,0.05;\tilde\beta)  = 3.18$.
Some values of  $m(n,\alpha)$  and  $t(n,\alpha)$ appear in Tables   \ref{Tab3} and \ref{critical values}.

\renewcommand{\thetable}{B.1}
\begin{table}[H]
\begin{centering}
\begin{tabular}{ccccc}
\multirow{3}{*}{$n$} & \multirow{3}{*}{$\alpha$} & \multicolumn{3}{c}{$S(n)$}\tabularnewline

 &  & 4 & 5 & 6\tabularnewline
\hline 
\multirow{2}{*}{50} & 10\% & 2.77 & 2.79 & 2.93\tabularnewline
 & 5\% & 2.89 & 3.14 & 3.43\tabularnewline
\multirow{2}{*}{100} & 10\% & 2.64 & 2.79 & 2.88\tabularnewline
 & 5\% & 2.92 & 3.14 & 3.30\tabularnewline
\multirow{2}{*}{150} & 10\% & 2.61 & 2.78 & 2.92\tabularnewline
 & 5\% & 2.96 & 3.07 & 3.20\tabularnewline
  \multirow{2}{*}{300} & 10\% & 2.78 &2.97 & 3.04\tabularnewline
 & 5\% &3.04 & 3.19 & 3.32\tabularnewline
  \multirow{2}{*}{500} & 10\% & 2.83 &2.93 & 3.05\tabularnewline
 & 5\% &3.04 & 3.18 & 3.29\tabularnewline
\end{tabular}
\par\end{centering}
\caption{\label{Tab3} Some critical values  $m(n,\alpha)$ of  $\mathcal{M}_{D(n)}$ in computing $\mathcal{P}_{{R}(\alpha)}$ for  $\mathbb{H}_{0} : \Phi(\cdot)$ (simple null hypothesis)}
\end{table}

\renewcommand{\thetable}{B.2}

\begin{table}[H]
\begin{centering}
\begin{tabular}{cccccc}
$\alpha/n $ &$50$& $100 $ & $ 150 $ &  $300 $ &  $500$  \\[1pt]
\hline 

$0.10$ & 2.15 &  2.33 & 2.42 & 2.57 & 2.67  \\
$0.05$ & 2.52 & 2.73 &  2.83& 3.00& 3.10   \\ [5pt]
\end{tabular}

\par\end{centering}
\caption{\label{critical values} Some critical values of test statistic $ \mathcal{T}_{n} $  for $\mathbb{H}$ : $\Phi\big(({x-\beta_{1})/}{\beta_{2}}\big)$ (unknown parameters)}
\end{table}

The computation of critical values  $c(n,\alpha)$ for $\mathcal{P}_{{R}(\alpha)}$ is rather straightforward and one can get a good approximation with 25~000 Monte Carlo replications. Such a number is required because this test statistic has a distribution with a discrete component. Table \ref{Tab5} lists some critical values $\tilde{c}(n,\alpha)$ for $\mathcal{P}_{\tilde{Q}(\alpha)}(\tilde\beta)$. \textcolor{black}{Using these leads to probabilities of a type 1 error very close to the nominal $5\%$ and $10\%$.}

\renewcommand{\thetable}{B.3}
\begin{table}[H]
\begin{centering}
\begin{tabular}{ccccc}
\multirow{3}{*}{$n$} & \multirow{3}{*}{$\alpha$} & \multicolumn{3}{c}{$S(n)$}\tabularnewline

 &  & 4 & 5 & 6\tabularnewline
\hline 
\multirow{2}{*}{50} & 10\% & 7.96 & 8.43 & 8.43\tabularnewline
 & 5\% & 10.48 & 10.79 & 10.86\tabularnewline
\multirow{2}{*}{100} & 10\% & 8.10 & 8.29 & 8.31\tabularnewline
 & 5\% & 10.43 & 10.67 & 10.70\tabularnewline
\multirow{2}{*}{150} & 10\% & 8.11 & 8.32 & 8.39\tabularnewline
 & 5\% & 10.33 & 10.46 & 10.57\tabularnewline
 \multirow{2}{*}{300} & 10\% & 7.88 &8.07 & 8.15\tabularnewline
 & 5\% &10.01 & 10.18 & 10.24\tabularnewline
  \multirow{2}{*}{500} & 10\% & 7.78 &7.94 & 7.95\tabularnewline
 & 5\% &9.71 & 9.88 & 9.92\tabularnewline
\end{tabular}
\par\end{centering}
\caption{\label{Tab5} Some critical values of $\tilde{c}(n,\alpha)$ for testing  $\mathbb{H}$ : $\Phi\big(({x-\beta_{1})/}{\beta_{2}}\big)$  (unknown parameters) with test statistic $\mathcal{P}_{\tilde{Q}(\alpha)}(\tilde\beta)$  }
\end{table}

One must  generally be careful when using GoF tests involving an oracle. Such a construction creates a null distribution which is a complex mixture of two components: one when the oracle accepts and another one when it rejects. Regarding statistic $\mathcal{P}_{{R}(\alpha)}$, this mixture distribution is steep enough so that no difficulty occurs in computing its $\alpha$-th critical values in the range of conditions we have investigated. However, the null CDF of  the counterpart $\mathcal{P}_{\tilde{R}(\alpha)}(\tilde\beta)$ of  $\mathcal{P}_{{R}(\alpha)}$ is  approximately  $1-\alpha$  for a large set of values. This creates instability in computing its $\alpha$-th critical value,  which must be resolved by using over two million MC replications, a serious defect of the procedure. The use here of $\mathcal{P}_{\tilde{Q}(\alpha)}(\tilde\beta)$, with the penalty $A(1.5;\tilde{\beta})$, slightly less than the Akaike $A(2;\tilde{\beta})$, provides a null distribution where the required critical values are easier to approximate :
 if necessary, these can be obtained with as little as 25~000 replications for $\alpha = 0.10$ and $0.05$. 

With a simple null hypothesis as in  Section 2, the power of the oracle $\mathcal{M}_{D(n)}$  increases with  $S(n)$ and this in turn affects favourably the  power  of  $\mathcal{P}_{{R}(\alpha)}$. As a consequence, we recommend using a large value, e.g. $S(n) = 6$, as in the simulations of Section 2.5 and the smiling baby data of Section 3.5.4. However, in the context of a composite null hypothesis, the oracle BCMR is not affected by this choice, and this reflects on the powers of $\mathcal{P}_{\tilde{Q}(\alpha)}(\tilde\beta)$ which are rather stable as a function of $S(n)$. Hence a small value can be used for the GoF test and  here we have taken  $S(n) = 4$.  However,  to extract useful insight from a B-plot, we recommend first computing the simultaneous acceptance regions with  $S(n) = 4$ and, if necessary, use  $S(n) = 6$  to get a  richer picture, as we have done in the examples of Section 3.5 and Appendix~A.

\vspace*{15pt}

%   ############    Appendix C : Simulation experiment Case 3   ################

\section{The simulation experiment for a composite Gaussian null hypothesis and related comments}

{In this appendix, we describe the setting and results of the simulation experiment discussed in Section 3.4. 
Note beforehand that all computations and simulations in the present paper, both in the previous sections and the present appendices, were performed using the Mathematica language
(Wolfram Research, Inc., Mathematica, Version 12.1, Champaign, IL, 2020)
and the random number generators in the program. \textcolor{black}{Note also that the calculation of test statistic $\mathcal{P}_{{R}(\alpha)}$  and  $\mathcal{P}_{\tilde{Q}(\alpha)}(\tilde\beta)$ is rather quick. For example, with $S(n) = 6$ and $n = 500$, our test statistics are computed in about 0.10 second on a MacBook Pro M2 running MacOS Ventura 13.2.1. Associated computational efforts are necessary in the computation of (\ref{eq:p-value case 3}); for example, the approximation of any $u(n, \alpha; \{ r,\ldots,s \}$ or $\ell(n, \alpha; \{ r,\ldots,s \}$  when  $S(n) = 4$ and $n = 100$,  based on a reasonnable 10~000 replications, requires less than two minutes. Thus the computational effort to use the statistics of the paper in practice can be considered marginal.}

We recall that  the null hypothesis is the composite $\ensuremath{{\mathbb{H}}:F(x)=\Phi\big((x-\beta_{1})/\beta_{2}\big)}$,
i.e. we consider testing GoF to the Gaussian distribution. We estimate $\beta$
by the MLE  $\tilde{\beta}=(\bar{X},S^{2})^\prime$, so that $\beta_1(F)$, $\beta_{2}^{2}(F)$ are the mean and variance of $F(\cdot)$. 

The alternatives were selected from some extensive simulation studies and chosen with care to cover a fair range of
shapes (see Figure \ref{A.1})  of $\textnormal{CC}(\cdot\; ;\beta(F))$ while embedding, either exactly or approximately, the Gaussian distribution. 
They are:
\begin{itemize}
\item $\mathbb{A}_{1}(\theta)$, the Tukey distributions with quantile function $\theta^{-1}(q^{\theta}-(1-q)^{\theta})$
if $\theta\neq0$ and $\log(q/(1-q)$) when $\theta=0$; these are symmetric about 0 unimodal distributions having support
$[-1/\theta,1/\theta]$, if $\theta>0$ and $\mathbb{R}$ otherwise;
$\mathbb{A}_{1}(0.14)$ is close to a $N(0,2.142)$, see Pearson, D'agostino \& Bowman (1977);
\item $\mathbb{A}_{2}(\theta)$, normal distributions perturbed by cosine functions with densities $\phi(x)\,[1+\theta\,\cos(4\pi\,\Phi(x))],$
$\theta\in[0,1]$ on $\mathbb{R}$ which for $\theta>0.3$ are visually clearly trimodal, see Inglot, Jurlewicz \& Ledwina (1990);
the case $\theta=0$ yields the $N(0,1)$;
\item $\mathbb{A}_{3}(\theta)$ = $\mathbb{A}_{3}^{0}(\theta)$, the two-piece normal distributions of Section 2.5; these asymmetric distributions have a left tail proportional to
a Gaussian, a fat right tail when $\theta>1$ and a short one otherwise  (see Experiment $C_{2}$ in  Boero, Smith \& Wallis (2004);  for some history about this distribution, see Wallis (2014));
\item $\mathbb{A}_{4}(\theta)$ = $\mathbb{A}_{4}^{0}(\theta)$, the Fan local model defined in Section 2.5;
these densities are asymmetric, bimodal and their tails coincide with
those of the $N(0,1)$ which is $\mathbb{A}_{4}(0)$ (see Example 5 in Fan, 1996); 
 \item $\mathbb{A}_{5}(\theta)$ = $\mathbb{A}_{5}^{0}(\theta)$, the normal contamination model defined in Section 2.5;
the case $\theta=0$ yields the $N(0,1)$; see Pearson, D'agostino \& Bowman (1977);
\item $\mathbb{A}_{6}(\theta)$ = $\mathbb{A}_{6}^{0}(\theta)$, Anderson's skewed distribution of Section 2.5
the case $\theta=0$ yields the $N(0,1)$ (see Experiment $C_{3}$ in Boero, Smith \& Wallis, (2004);
\item $\mathbb{A}_{7}(\theta)$ = $\mathbb{A}_{7}^{0}(\theta)$, the Mason \& Schuenemeyer (1983) symmetric about zero distribution with CDF $J(\Phi(x),0.15, \theta)$; the case $\theta = 0$ yields the $N(0, 1)$ (see \'Cmiel,  Inglot \& Ledwina, 2020);
\item $\mathbb{A}_{8}(\theta)$, Johnson's SU distributions, with $X=\sinh(Z/\theta)$,
$\theta>0$ and $Z\sim N(0,1)$, yielding symmetric about zero and
unimodal densities;  $\mathbb{A}_{8}(3.5)$ is approximately  $N(0,0.22)$, see Pearson, D'agostino \& Bowman (1977); 
\item $\mathbb{A}_{9}(\theta)$ = $\mathbb{A}_{9}^{0}(\theta)$, the Lehmann contamination model from Section 2.5,
a contamination model skewed to the left with $\mathbb{A}_{9}(0)=N(0,1)$ (see \'Cmiel, Inglot \& Ledwina, 2020);
\item $\mathbb{A}_{10}(\theta)$, the Lehmann model with CDF  $(\Phi(x))^\theta$;  of course the case $\theta = 1$ yields the $N(0,1)$ and here we take $0<\theta \leq 1$;
\item $\mathbb{A}_{11}(\theta)$, the generalized error distribution (GED) with density proportional to $\textrm{exp}(-|x|^\theta/\theta)$; the case $\theta = 2$ yields the $N(0,1)$ and here we consider $\theta \leq 2$;
\item $\mathbb{A}_{12}(\theta)$, a symmetric Pareto contamination model with CDF  given by $(1-\theta)N(0,1)+\theta \;\Pi(5)$, where $ \Pi(\delta)$  is the symmetric Pareto distribution (see model $\mathbb{M}_7$  in \'Cmiel, Inglot \& Ledwina, 2020).
\end{itemize}

Additionally, alternative $\mathbb{A}_{8}^{0}(\theta)$ in the simulation of Section 2.5 has been extensively used in Experiment $D_{2}$ of Boero, Smith \& Wallis  (2004).

As competitors to $\mathcal{P}_{\tilde{Q}(\alpha)}(\widetilde{\beta})$,
we have considered the following tests:
\begin{itemize}
\item the Anderson-Darling (AD) test adjusted for unknown parameters; 
\item the Shapiro-Wilks (SW) test;
\item the del Barrio, Cuesta-Albertos, Matran \& Rodriguez  (1999)  test BCMR.
\item at the reviewer's request, the data driven smooth test of Janic \& Ledwina (2009) with test statistic $W^{*}_{S1}(\tilde{\beta}[ns])$.
\end{itemize}

We recall that our main focus here pertains to the interpretation of the new components,  provided the related test statistic $\mathcal{P}_{\tilde{Q}(\alpha)}(\widetilde{\beta})$ exhibits a generally good behavior, as we now show.

\renewcommand{\thefigure}{C.1}

\begin{figure}[H]
\begin{centering}
\includegraphics[width = 1.0\linewidth]{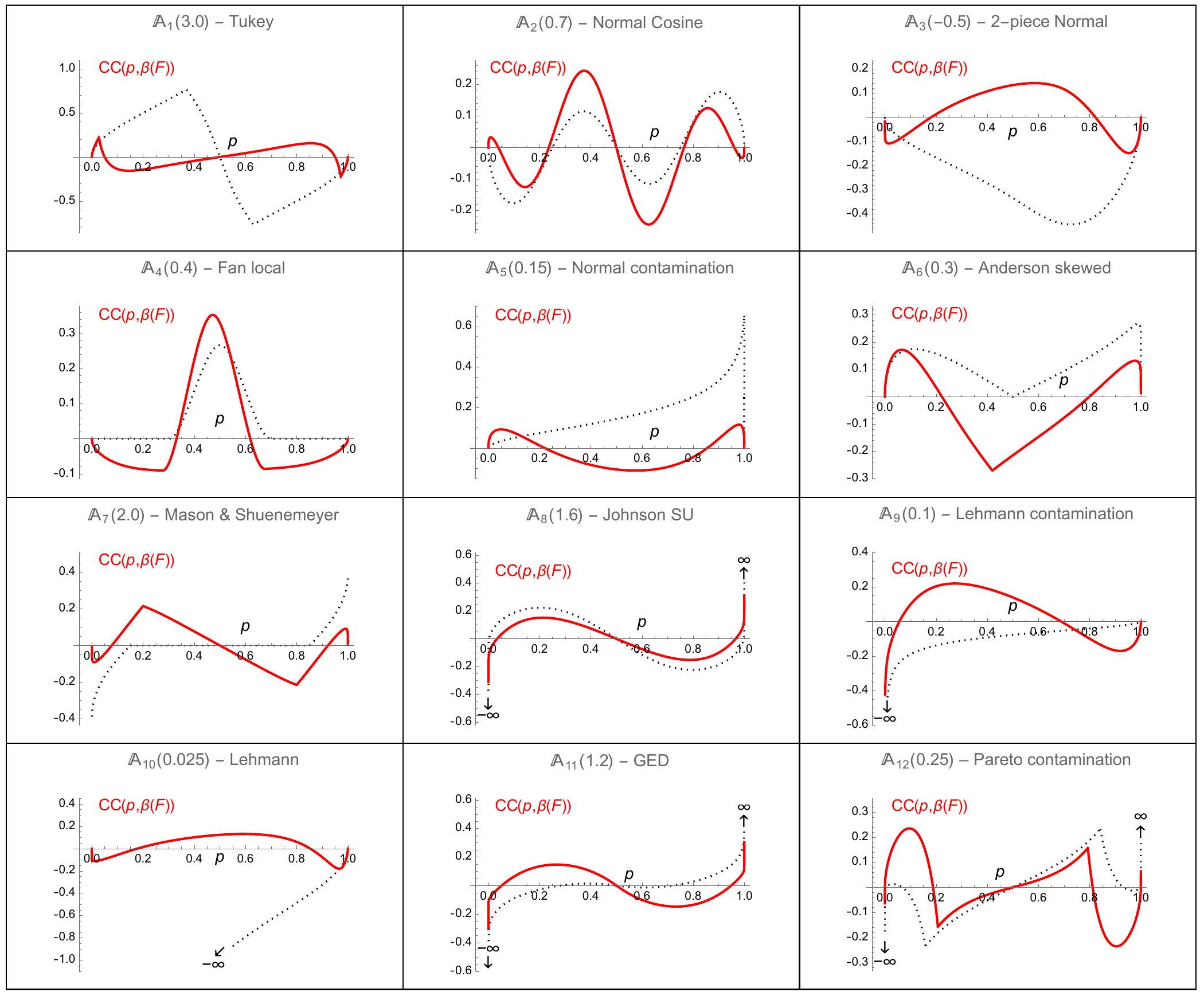}
\caption{\label{A.1} $\textnormal{CC}(\cdot;\beta(F))$ (solid red) for
the alternative distributions in Table~\ref{tab:Power-at-level 5=000025 Case 3} for testing the composite null hypothesis 
$\Phi\big((x-\beta_{1})/\beta_{2}\big) $ with $(\beta_{1},~\beta_{2})$ unknown. The black dotted curve represents  $\textnormal{CC}(\cdot)$ of (2.1) corresponding to the simple null model 
$\Phi(x) $.}
\par\end{centering}
\end{figure}

Taking $\alpha=0.05$ and $n=100$, we have investigated $\mathcal{P}_{\tilde{Q}(\alpha)}(\widetilde{\beta})$
with the del Barrio, Cuesta-Albertos, Matran \& Rodriguez (1999) BCMR oracle test using $S(n)=4$; this yields $a(n,0.05;\tilde\beta)  = 3.18$, see Appendix~B. The power functions
for the twelve alternatives were simulated in their $\theta$ range
for each of the tests $\mathcal{M}_{D(n)}(\tilde{\beta})$ of (15) with $D(n) = 31$ and $D(n) = 127$, AD, SW,  BCMR, $W^{*}_{S1}(\tilde{\beta}[ns])$ and $\mathcal{P}_{\tilde{Q}(\alpha)}(\widetilde{\beta})$.
The 5\% critical value for each test was obtained from 100~000 replications
under the null distribution, while the powers were computed from 10~000
Monte Carlo runs. As in Section 2.5,
we extracted from each power curve one representative value of $\theta$
which provided interesting powers. As a result of these choices, the
obtained powers give a comprehensive view of the comparative behaviour
of the above tests in a wide range of situations. 
The results are reported in Table~\ref{tab:Power-at-level 5=000025 Case 3} along the selected value of $\theta$ and
roughly sorted from thin to fat-tailed.
As can be seen,  none of the other tests dominates $\mathcal{P}_{\tilde{Q}(\alpha)}(\widetilde{\beta})$, which emerges as a good competitor over our range of alternatives.

\renewcommand{\thetable}{C.2}

\begin{table}
\begin{centering}
\begin{tabular}{ccccccccc}
$Alternative$ &$\mathcal{M}_{31}(\tilde{\beta})$ &$\mathcal{M}_{127}(\tilde{\beta})$ & AD & SW & BCMR&  $W^{*}_{S1}(\tilde{\beta}[ns])$ & $\mathcal{P}_{\tilde{Q}(\alpha)}(\widetilde{\beta})$ \\[3pt]
\hline 
$\mathbb{A}_{1}(3.0)$ & 42 &38& 47 & 74 & 70 & 59 & 68  \\
$\mathbb{A}_{2}(0.7)$ & 64 &60& 75 & 40 & 41 & 57 &  57  \\
$\mathbb{A}_{3}(-0.5)$ & 27 &29& 43 & 46 & 47& 46 & 45   \\
$\mathbb{A}_{4}(0.4)$  & 82 &80& 78 & 43 & 45& 39 & 65  \\
$\mathbb{A}_{5}(0.15)$ &17 &21& 28 & 29 &  30 & 26 &28   \\
$\mathbb{A}_{6}(0.3)$& 58 & 58& 73 & 68 & 70& 64 & 69   \\
$\mathbb{A}_{7}(2.0)$ & 47&  44 &56 & 47 & 51& 60 & 56   \\
$\mathbb{A}_{8}(1.6)$& 34 &34 & 47 & 55 & 58& 61 &  53   \\
$\mathbb{A}_{9}(0.1)$& 53 &55& 68 & 80 & 82 & 80 & 75   \\ 
$\mathbb{A}_{10}(0.025)$& 27 &28& 43 & 57 & 55 & 47 & 50   \\
$\mathbb{A}_{11}(1.2)$& 39 &37& 54 & 54 & 56 & 60 & 56   \\
$\mathbb{A}_{12}(0.25)$&77 &74& 71 & 58 & 56 & 63 & 66   \\[5pt]
\end{tabular}
\par\end{centering}
\caption{\label{tab:Power-at-level 5=000025 Case 3} Powers ($n=100$, $\alpha=0.05$) of  $\mathcal{M}_{D(n)}(\tilde{\beta})$ of (15) with $D(n) = 31$ and $D(n) = 127$, the  Anderson-Darling (AD),  the Shapiro-Wilks (SW), the oracle BCMR, the Janic\&Ledwina (2009) test ($W^{*}_{S1}(\tilde{\beta}[ns])$) and  our $\mathcal{P}_{\tilde{Q}(\alpha)}(\widetilde{\beta})$  tests
for $\mathbb{H}$: $\Phi\big(({x-\beta_{1})/}{\beta_{2}}\big)$ with $(\beta_{1},~\beta_{2})$ unknown, 
against the set of alternatives $\mathbb{A}_{1}(\theta)$ to  $\mathbb{A}_{12}(\theta)$.
}
\end{table}

%The tests of the present paper are based on $CC(\cdot)$, which are aggregated
%Fourier coefficients (FC) of the respective comparison density.
%For  alternatives defined as in Section 2.5 and in the present section, a plot of $CC(\cdot)$ can be produced, and given
%a variety of such plots, they can be scrutinized to eliminate alternatives
%with similar shape. Moreover, others can be consciously sought to
%add diversity in any desired way. Because the magnitude of the FC
%has a direct relationship with the behaviour of our tests, this helps
%in producing a less bias appreciation of the comparative power of
%our proposal. 

Power comparisons between the case where $F_{0}(\cdot\,;\,\beta)$ is
fully specified and those where $\beta$ is estimated seldom appear
in the literature. Here, in addition to our concern regarding well-balanced
set of alternatives, we have made the deliberate choice of selecting
three alternatives in common in Tables~1
and \ref{tab:Power-at-level 5=000025 Case 3}, namely 
the Fan local alternative  ($\mathbb{A}_{4}^{0}$ and $\mathbb{A}_{4}$), 
the normal contamination ($\mathbb{A}_{5}^{0}$ and $\mathbb{A}_{5}$)
and the Lehmann contamination ($\mathbb{A}_{9}^{0}$ and $\mathbb{A}_{9}$).
This was done to explore
the difficulties in going from a simple to a composite null model.
It is interesting to see the strong impact the estimation process
has on the $\textnormal{CC}(\cdot)$. Basically, the expression for $\textnormal{CC}(\cdot)$
is modified from (1) to (11)
in which the denominator is different  while additionally, in (11), $\beta(F)$ plays a role. More precisely, in the case
of a simple null hypothesis, $\Phi(\cdot)$ is the reference distribution to an alternative $F(\cdot)$. When the null
hypothesis is composite, the reference CDF for the same $F(\cdot)$ is $\Phi\big(({x-\beta_{1}(F))/}{\beta_{2}(F)}\big)$ which
is now adjusted to $F(\cdot)$. The new denominator in (11) plays a strong role in the central region
of $(0, 1)$ while the standardization of $\Phi(\cdot)$ via the use of $\beta(F)$ affects the tails of the reference distribution
and thus plays an essential role for $p$ close to 0 and 1.

For the three alternatives in common,  Figure \ref{A.1} shows that pairs of $\textnormal{CC}(\cdot)$ and  $\textnormal{CC}(\cdot ; \,
\beta(F))$ can
be very different and this affects the powers in a way that is difficult to predict without such additional insight. Figure C.1 exhibits more cases where $\textnormal{CC}(\cdot)$ and the associated  $\textnormal{CC}(\cdot ; \,
\beta(F))$ are rather different, e.g. $\mathbb{A}_{1}$, $\mathbb{A}_{3}$, $\mathbb{A}_{7}$ and $\mathbb{A}_{10}$, with the selected parameter $\theta$. In particular, it can be noticed that the estimation of $\beta$ strongly affects the related shape of $\textnormal{CC}(\cdot ;\, \beta(F))$ in the case of relatively heavy-tailed alternatives, as then the 
pertaining $\beta_2(F)$ is often large. It should also be remembered that the required $\sqrt{n}$-consistency of the MLE of $\beta$ implies a substantial restriction on the allowable class
of alternatives $F(\cdot)$ to those having finite fourth moment.
Also, it should be stated that large differences
in the forms of the  \textnormal{CC}  are sometimes almost imperceptible at the level of densities.  This is the case for the Fan alternative  ($\mathbb{A}_{4}^{0}$(0.4) and $\mathbb{A}_{4}(0.4)$). Finally, going from
the simple null hypothesis to a composite one, the selection rule had to be adapted in an intricate way to preserve the good properties of our procedures. 

Hence, it can be concluded that the case of unknown parameters is a problem
whose complexity is of an order of magnitude above the fully specified case.

%   ############    Appendix D: Proof of propositions   ################

\section*{Appendix~D.~~Proofs}  \label{subsec:Proof prop 1}

\renewcommand{\theequation}{D. \arabic{equation}} % This line ads "Eq." in front of your equation numbering.

\subsection*{D.1.~~Proof of Proposition 1 }  We have that $\ensuremath{\hat{F}_{n}\bigl(F_{0}^{-1}(p_{s,j})\bigr)=n^{-1}\sum_{i=1}^{n}I(F_{0}(X_{i})\leq p_{s,j})}$
where $\ensuremath{F_{0}(X_{i})}$ are i.i.d. $\ensuremath{U(0,1)}$
under $\ensuremath{\mathbb{H}_{0}}$. Let $\ensuremath{\alpha_{n}(t),\;t\in[0,1],}$
denote the uniform empirical process. Then it holds that
\begin{align*}
\mathcal{M}_{D(n)} &= \max_{1\leq j\leq D(n)}\frac{\left|\alpha_{n}(p_{S(n),j})\right|}{(p_{S(n),j}(1-p_{S(n),j}))^{1/2}}.
\end{align*}
Because $\ensuremath{D(n)=o(n^{2\delta}),\;\delta\in(0,1/2)}$, for
sufficiently large $\ensuremath{n}$ we get $\ensuremath{p_{S(n),1}\geq\epsilon_{n}=[\log n]^{3}/n}$.
Moreover, for the quantity
\begin{align*}
\sup_{\epsilon_{n}\leq t\leq1-\epsilon_{n}}\frac{|\alpha_{n}(t)|}{(t(1-t))^{1/2}},
\end{align*}
the Darling-Erd\"{o}s theorem holds, see Jaeschke (1979). This implies
that 
\begin{align}
\mathcal{M}_{D(n)} & =O_{P}((\log\log n)^{1/2}).\label{eq: order of m_D(n)}
\end{align}
Let $\ensuremath{c(n,\alpha)}$ denote the $\alpha-$th critical value
of $\mathcal{P}_{R(\alpha)}$. 
On the set $\ensuremath{\{\mathcal{M}_{D(n)}>m(n,\alpha)\}}$,
the value of $\mathcal{P}_{R(\alpha)}$ is equal to $\mathcal{P}_{D(n)}$.
Moreover, it holds that 
\begin{align*}
\textrm{pr}\big(\mathcal{P}_{ R(\alpha)} \geq c(n,\alpha)\big) &=\textrm{pr}\big(\mathcal{P}_{ R(\alpha)} \geq  c(n,\alpha)~|~\mathcal{M}_{D(n)} \leq m(n,\alpha)\big)\\
& \times \textrm{pr}\big({\mathcal M}_{D(n)} \leq m(n,\alpha)\big)+\;\textrm{pr}\big(\mathcal{P}_{D(n)} \geq  c(n,\alpha)\big)\\
&-\textrm{pr}\big(\mathcal{P}_{D(n)} \geq  c(n,\alpha), {\mathcal M}_{D(n)} \leq m(n,\alpha)\big).
\end{align*}
Hence, it is enough to show that, under $F(\cdot)$, the test rejecting
for large values of $\ensuremath{\mathcal{M}_{D(n)}}$ is consistent
and $ \textrm{pr}(\mathcal{P}_{D(n)}\geq c(n,\alpha))\rightarrow$ 1
as $\ensuremath{n\rightarrow\infty}$.

We start by showing that, under $F(\cdot)$, the test rejecting
for large values of $\ensuremath{\mathcal{M}_{D(n)}}$ is consistent. By (\ref{eq: order of m_D(n)}), $m(n,\alpha)$ cannot grow faster
than $\ensuremath{O((\log\log n)^{1/2})}$. On the other hand, using
the definition of $\ensuremath{(s_{0},j_{0})}$, we have for $n$ large enough
\begin{align}
 \textrm{pr}\big(\mathcal{M}_{D(n)}>m(n,\alpha)\big) & =  \textrm{pr}\big(\max_{1\leq j\leq D(n)}|n^{1/2}\hat{\gamma}_{j}(p_{s,j})|>m(n,\alpha)\big)  \label{eq:majoration for m_D(n)} \\
&   \geq\textrm{pr}\big(n^{1/2}|\hat{\gamma}_{j_{0}}(p_{s_{0},j_{0}})|>m(n,\alpha)\big) \label{eq:majoration for m_D(n)} \nonumber  \\
& =\textrm{pr}\Bigl(\Bigl|n^{1/2}\gamma_{j_{0}}(p_{s_{0},j_{0}})-V_{n}\Bigr|>m(n,\alpha)\Bigr),\nonumber 
\end{align}
where
\begin{align*}
V_{n} & =\frac{n^{1/2}[\hat{F}_{n}\bigl(F_{0}^{-1}(p_{s_{0},j_{0}})\bigr)-F\bigl(F_{0}^{-1}(p_{s_{0},j_{0}})\bigr)]}{\{p_{s_{0},j_{0}}(1-p_{s_{0},j_{0}})\}^{1/2}},
\end{align*}
while $\ensuremath{\gamma_{j_{0}}(p_{s_{0},j_{0}})=\gamma_{s_{0},j_{0}}}$
is defined in (4). The numerator in the
formula for $\ensuremath{V_{n}}$ is $\ensuremath{O_{P}(1)}$ while
the denominator's impact onto $\ensuremath{V_{n}}$ is at most of
the order $\ensuremath{(D(n))^{-1/2}}.$ Due to the assumption on
$\ensuremath{D(n)}$, the term $\ensuremath{V_{n}}$ is $\ensuremath{o_{P}(n^{\delta})}$, $\delta \in (0,1/2)$.
Hence, in view of (D.2) and the range
of $\ensuremath{m(n,\alpha)}$, we conclude that under $F(\cdot)$, $\ensuremath{\textrm{pr}\big(\mathcal{M}_{D(n)} > m(n,\alpha)\big)\to1}$.

Now we show convergence of $\textrm{pr}\big(\mathcal{P}_{D(n)} \geq  c(n,\alpha)\big)$ to 1. Because $\mathcal{P}_{R(\alpha)} \leq \mathcal{P}_{D(n)}\leq D(n)\times O_{P}(\log\log n)$, then
$\ensuremath{c(n,\alpha)=o(n^{2\delta}\times\log\log n)}$. Similarly
as in the case of $\ensuremath{\mathcal{M}_{D(n)}}$, we can write 
\begin{alignat*}{1}
\textrm{pr}\big(\mathcal{P}_{D(n)}\geq c(n,\alpha)\big) & \geq \textrm{pr}\big(n^{1/2}\left|\hat{\gamma}_{j_{0}}(p_{s_{0},j_{0}})\right| \geq c(n,\alpha)^{1/2}\big)\\
 & =\textrm{pr}\big(\left| n^{1/2} \gamma_{j_{0}}(p_{s_{0},j_{0}})-V_{n}\right|\geq c(n,\alpha)^{1/2}\big).
\end{alignat*}
Taking into account the rate of growth of $\ensuremath{c(n,\alpha)}$,
the same argument as above finishes the proof. $\hfill\ensuremath{\Box}$

\vspace*{20pt}

\subsection*{D.2.~~Proof of Proposition 2}  We start by reducing the consistency problem.  Let $\tilde c(n,\alpha)$ denote the $\alpha-$th critical value of ${\mathcal P}_{\tilde Q(\alpha)}(\tilde \beta)$. In the sequel, all probabilities are computed under $F(\cdot)$. Then,
\begin{align*}
\textrm{pr}\big({\mathcal P}_{\tilde Q(\alpha)}(\tilde \beta) \geq \tilde c(n,\alpha)\big) &=\textrm{pr}\big({\mathcal P}_{\tilde Q(\alpha)}(\tilde \beta) \geq \tilde c(n,\alpha)~|~{\mathcal T}_n \leq t(n,\alpha)\big)\\
& \times \textrm{pr}\big({\mathcal T}_n \leq t(n,\alpha)\big) \\
& +  \,\textrm{pr}\big({\mathcal P}_{\tilde Q(\alpha)}(\tilde \beta) \geq \tilde c(n,\alpha)~|~{\mathcal T}_n > t(n,\alpha)\big) \\
& \times\textrm{pr}\big({\mathcal T}_n > t(n,\alpha)\big).
\end{align*}
From pp.1-2 of the Supplementary Material to Ledwina and Wy{\l}upek (2015),  ${\mathcal T}_n$ is consistent under the assumption that $F(\cdot)$ possesses a finite second moment. Hence
$\textrm{pr}\big({\mathcal T}_n > t(n,\alpha)\big) \to 1$.
In view of the above, to prove that $ \textrm{pr}\big({\mathcal P}_{\tilde Q(\alpha)}(\tilde \beta) \geq \tilde c(n,\alpha)\big) \to 1$, it is enough to show that
\begin{align}  
\textrm{pr}\big({\mathcal P}_{\tilde Q(\alpha)}(\tilde \beta) < \tilde c(n,\alpha)~|~{\mathcal T}_n > t(n,\alpha)\big) \to 0. 
\end{align}
The next step is to get the rate of growth of $\tilde c(n,\alpha)$. ${\mathcal P}_{\tilde Q(\alpha)}(\tilde \beta) \leq {\mathcal P}_{D(n)}(\tilde \beta) \leq U_n,$ where
$$
U_n=D(n)\big[n \sup_{0\leq p \leq 1}|p-\hat F_n(\tilde \beta_2F_0^{-1}(p)+\tilde \beta_1)|^2\big]\times \big[\min_{1 \leq j \leq D(n)}\sigma_{S(n),j}^2\big]^{-1}.
$$
Under $\mathbb H$, the first expression in squared brackets is $O_P(1)$  by Durbin's (1973) theorem. The second expression is $O(D(n))$. As a consequence, $U_n=O_P(D^2(n))$ and $\tilde c(n,\alpha)$ does not grow faster than $D^2(n)$. Hence, in view of the assumption on $D(n)$,  $\tilde c(n,\alpha)=o(n).$\\

Recall from Section 3.2 the indices  $s_0$ and $j_0$  such that $\textnormal{CC}(p_{s_0,j_0};\beta(F))\neq 0.$
We have from (13), (14)
$$
{\mathcal P}_{d(s_0)}(\tilde \beta)  = n \sum_{j=1}^{d(s_0)} \Bigl[ \frac{p_{s_0,j}- \hat F_n\bigl(\tilde \beta_2 F_0^{-1}(p_{s_0,j})+\tilde \beta_1\bigr)}{(p_{s_0,j}(1-p_{s_0,j}))^{1/2}}\Bigr]^2 =
n \sum_{j=1}^{d(s_0)} \Bigl[\widehat{\textnormal{CC}}(p_{s_0,j};\tilde \beta)\Bigr]^2.
$$ 
Now, we show that 
\begin{align}
\textrm{pr}\big(n^{1/2} |\widehat{\textnormal{CC}}(p_{s_0,j_0};\tilde \beta)| \geq  (\tilde c(n,\alpha))^{1/2}\big) \to 1,\quad n \to \infty. 
\end{align} \label{D.4}
To this end, observe that
\begin{align} \label{D.5}
n^{1/2}\;\widehat{\textnormal{CC}}(p_{s_0,j_0};\tilde \beta) & =n^{1/2}\;{\textnormal{CC}}(p_{s_0,j_0}; \beta(F))  \\
&+ \{W_n^{(1)}(s_0,j_0;\tilde \beta)+W_n^{(2)}(s_0,j_0;\tilde \beta)\} \times [\sigma_{s_0,j_0}]^{-1},  \nonumber 
\end{align}
where
\begin{align}
W_n^{(1)}(s_0,j_0;\tilde \beta)=n^{1/2} \{F(\tilde \beta_2 F_0^{-1}(p_{s_0,j_0})+\tilde \beta_1)-\hat F_n(\tilde \beta_2 F_0^{-1}(p_{s_0,j_0})+\tilde \beta_1)\},\label{D.6}
\end{align}
while
\begin{align}
W_n^{(2)}(s_0,j_0;\tilde \beta)= n^{1/2} \{F(\beta_2(F)F_0^{-1}(p_{s_0,j_0})+\beta_1(F)) - F(\tilde \beta_2 F_0^{-1}(p_{s_0,j_0})+\tilde \beta_1)\}. \label{D.7}
\end{align}

The deterministic term in (\ref{D.6}) is $O(n^{1/2})$. The term $W_n^{(1)}(s_0,j_0;\tilde \beta)$ can be majorized by $n^{1/2} \sup_{x \in \mathbb R}|F(x)-\hat F_n(x)|$
and is thus $O_P(1)$. Using the equality $F(x)-F(y)=(x-y)f(z^*)$, where $f(\cdot)$ is the density of $F(\cdot)$ and $\min\{x,y\} \leq z^* \leq \max\{x,y\}$, we see that 
\begin{align}
W_n^{(2)}(s_0,j_0;\tilde \beta)= n^{1/2} \{(\tilde \beta_2 - \beta_2(F))F_0^{-1}(p_{s_0,j_0}) + (\tilde \beta_1 - \beta_1(F))\}f(Z^*). \label{D.8}
\end{align}
Under the assumption on the fourth moment of $F(\cdot)$, $\tilde \beta_1$ and $\tilde \beta_2$ are $n^{1/2}$-consistent. By boundedness of $f(\cdot)$, the expression (\ref{D.7}) is $O_P(1)$.  Thus, by the above, $n^{1/2} \,\widehat{\textnormal{CC}}(p_{s_0,j_0};\tilde \beta)=O_P(n^{1/2})$. Because $\tilde c(n,\alpha)=o(n)$, we get  ${\mathcal P}_{d(s_0)}(\tilde \beta) \to \infty$.\\
Now, 
\begin{gather} 
\textrm{pr}\big(\tilde Q(\alpha) < d(s_0), {\mathcal T}_n > t(n,\alpha)\big)  \nonumber \\
\leq  \sum_{s=1}^{s_0 - 1}\textrm{pr}\big({\mathcal P}_{d(s)}(\tilde \beta) - 1.5\times d(s) \geq {\mathcal P}_{d(s_0)}(\tilde \beta) -1.5 \times d(s_0)\big). \label{D.9}
\end{gather}

For $s < s_0$ it holds that  $\textnormal{CC}(p_{s,j};\beta(F)) =0,  (j=1,\ldots,d(s)).$ Therefore, by (\ref{D.5}) to (\ref{D.8}) applied to such $s$ and related $p_{s,j}$, we have ${\mathcal P}_{d(s)}(\tilde \beta)= O_P(1)$. But it was earlier shown that ${\mathcal P}_{d(s_0)}(\tilde \beta) \to \infty$  as $n \to \infty$.  It follows that  $\textrm{pr}\big(\tilde Q(\alpha) < d(s_0)\big) \to 0$ on the set $\{{\mathcal T_{n}} > t(n,\alpha)\} $, as $n \to \infty$.\\

Getting back to (D.3), we have
\begin{gather} \label{D.10}
\textrm{pr}\big({\mathcal P}_{\tilde Q(\alpha)}(\tilde \beta) < \tilde{c}(n,\alpha)\big) = \textrm{pr}\big({\mathcal P}_{\tilde Q(\alpha)}(\tilde \beta) < \tilde{c}(n,\alpha), \tilde Q(\alpha) < d(s_0)\big) \\
+\,\sum_{s=s_0}^{D(n)}\textrm{pr}\big({\mathcal P}_{\tilde Q(\alpha)}(\tilde \beta) < \tilde{c}(n,\alpha), \tilde Q(\alpha) = d(s)\big). \nonumber
\end{gather} 
If $\tilde Q(\alpha) \geq d(s_0)$, then ${\mathcal P}_{\tilde Q(\alpha)}(\tilde \beta) \geq {\mathcal P}_{d(s_0)}(\tilde \beta)$. 
Moreover, on the set $\{{\mathcal T_{n}} > t(n,\alpha)\}$, the first summand in (\ref{D.10}) is $o(1)$. By the above
\begin{gather}
\textrm{pr}\big({\mathcal P}_{\tilde Q(\alpha)}(\tilde \beta) < \tilde{c}(n,\alpha)\big) \leq o(1) + D(n)~\textrm{pr}\big({\mathcal P}_{d(s_0)}(\tilde \beta) < \tilde{c}(n,\alpha)\big) \nonumber \\
\leq o(1) + D(n)~\textrm{pr}\big(n^{1/2} |\widehat{\textnormal{CC}}(p_{s_0,j_0};\tilde \beta)| <  (\tilde c(n,\alpha))^{1/2}\big).  \label{D.11}
\end{gather} 

This shows that we need to sharpen (D.4) by studying the rate at which the probability appearing in  (\ref{D.11}) tends to 0. But in view of (\ref{D.5})--(\ref{D.8}) the event ${\mathbb E}_n=\{n^{1/2} \, |\widehat{\textnormal{CC}}(p_{s_0,j_0};\tilde \beta)| <  (\tilde{c}(n,\alpha))^{1/2} \}$ reads as
\begin{align*}
{\mathbb E}_n= \{n^{1/2}\, l_n < W_n^{(1)}(s_0,j_0;\tilde \beta)+W_n^{(2)}(s_0,j_0;\tilde \beta) < n^{1/2} \, u_n\},
\end{align*}
where
$$
l_n=\sigma_{s_0,j_0}\Bigl\{-(n^{-1}\tilde c(n,\alpha))^{1/2}- \textnormal{CC}(p_{s_0,j_0};\beta(F)) \Bigr\},\\
$$
$$
u_n =\sigma_{s_0,j_0}\Bigl\{+(n^{-1}\tilde c(n,\alpha))^{1/2}- \textnormal{CC}(p_{s_0,j_0};\beta(F)) \Bigr\}\\.
$$
Because $\tilde c(n,\alpha)=o(n)$, we get $l_n=O(1)$ and $u_n=O(1)$. If $\textnormal{CC}(p_{s_0,j_0};\beta(F)) < 0$, then we can write 
$\textrm{pr}({\mathbb E}_n) \leq \textrm{pr} \big(W_n^{(1)}(s_0,j_0;\tilde \beta)+W_n^{(2)}(s_0,j_0;\tilde \beta) > n^{1/2}\, l_n\big)$. Otherwise, we can consider 
$\textrm{pr}\big({\mathbb E}_n) \leq \textrm{pr} (-W_n^{(1)}(s_0,j_0;\tilde \beta)+W_n^{(2)}(s_0,j_0;\tilde \beta)> n^{1/2} (-u_n)\big)$. Hence, the triangle inequality, the DKW inequality applied to $W_n^{(1)}(s_0,j_0;\tilde \beta)$ and Markov's inequality applied to both terms of $W_n^{(2)}(s_0,j_0;\tilde \beta)$ appearing in (\ref{D.8}) show that $\textrm{pr}({\mathbb E}_n)=O(n^{-1})$. In view of the assumption $D(n)=o(n^{1/2})$ we have $D(n)\,\textrm{pr}({\mathbb E}_n)=o(1)$ and by (\ref{D.11}), the proof is complete. \hfill{$\Box$}
\vspace*{20pt}

\subsection*{D.3.~~Proof of Remark 2} The relations  (\ref{D.5})--(\ref{D.8}), expressed in terms of an arbitrary $p \in [\epsilon, 1-\epsilon],$ imply that 
$$
\sup_{\epsilon \leq p \leq 1- \epsilon} n^{1/2} |\widehat{\textnormal{CC}}(p;\tilde \beta)-\textnormal{CC}(p;\beta(F))|=O_P(1).
$$
Hence the statement of Remark 2 follows.\hfill{$\Box$}
\vspace{1cm}

\renewcommand\refname{References for Appendices}

\end{document}